\documentclass[sigconf]{acmart}

\AtBeginDocument{%
  }

\copyrightyear{2024}
\acmYear{2024}
\setcopyright{acmlicensed}\acmConference[CIKM '24]{Proceedings of the 33rd ACM International Conference on Information and Knowledge Management}{October 21--25, 2024}{Boise, ID, USA}
\acmBooktitle{Proceedings of the 33rd ACM International Conference on Information and Knowledge Management (CIKM '24), October 21--25, 2024, Boise, ID, USA}
\acmDOI{10.1145/3627673.3679701}
\acmISBN{979-8-4007-0436-9/24/10}

\usepackage{xcolor}
\usepackage{hyperref}
\usepackage{algorithm,algorithmic}
\usepackage{bm}
\usepackage{multirow}
\usepackage{subfig}
\usepackage{enumitem}
\usepackage{soul}
\usepackage{balance}
\graphicspath{{./Fig/}}

\begin{document}

\title{HGCH: A Hyperbolic Graph Convolution Network Model for Heterogeneous Collaborative Graph Recommendation}

\author{Lu Zhang}
\email{luzhang_cs@hust.edu.cn}
\orcid{0000-0003-0225-7398}
\affiliation{%
  \institution{Huazhong University of Science and Technology}
  \city{Wuhan}
  \country{China}
}

\author{Ning Wu}
\authornote{Corresponding Author}
\email{wuning@buaa.edu.cn}
\orcid{0000-0001-8526-1581}
\affiliation{%
  \institution{Beihang University}
  \city{Beijing}
  \country{China}
}

\renewcommand{\shortauthors}{Lu Zhang \& Ning Wu}

\begin{abstract}
User-item interaction data in collaborative filtering and graph modeling tasks often exhibit power-law characteristics, which suggest the suitability of hyperbolic space modeling. Hyperbolic Graph Convolution Neural Networks (HGCNs) are a novel technique that leverages the advantages of GCN and hyperbolic space, and then achieves remarkable results. However, existing HGCN methods have several drawbacks: they fail to fully leverage hyperbolic space properties due to arbitrary embedding initialization and imprecise tangent space aggregation; they overlook auxiliary information that could enrich the collaborative graph; and their training convergence is slow due to margin ranking loss and random negative sampling. To overcome these challenges, we propose Hyperbolic Graph Collaborative for Heterogeneous Recommendation (HGCH), an enhanced HGCN-based model for collaborative filtering that integrates diverse side information into a heterogeneous collaborative graph and improves training convergence speed. HGCH first preserves the long-tailed nature of the graph by initializing node embeddings with power law prior; then it aggregates neighbors in hyperbolic space using the gyromidpoint method for accurate computation; finally, it fuses multiple embeddings from different hyperbolic spaces by the gate fusion with prior. Moreover, HGCH employs a hyperbolic user-specific negative sampling to speed up convergence. We evaluate HGCH on four real datasets, and the results show that HGCH achieves competitive results and outperforms leading baselines, including HGCNs. Extensive ablation studies further confirm its effectiveness.
\end{abstract}

\ccsdesc[500]{Information systems~Recommender systems}
\ccsdesc[500]{Information systems~Collaborative filtering}
\ccsdesc[500]{Computing methodologies~Neural networks}

\keywords{Recommender Systems; Hyperbolic Embeddings; Graph Convolutions}


\maketitle

\section{Introduction}

Recommender systems play a vital role in addressing information overload by offering personalized services to users and enhancing business revenue. One of the core methods employed in these systems is Collaborative Filtering (CF), which predicts user preferences based on historical user-item interactions. CF has been a focal point of research, leading to significant advancements.

Recent CF approaches leverage hyperbolic space, effectively capturing the power-law distribution and hierarchical structure in user-item interaction data. Prior studies \cite{TLS1,TLS2,TLS3} have revealed that such data often exhibit a tree-like structure, where the number of child nodes grows exponentially with distance from the root. This structure is distorted when modeled in Euclidean space, which has a polynomial growth of space capacity. In contrast, hyperbolic space, with its negative curvature, supports exponential growth of space capacity, making it an ideal fit for modeling tree-like data. Recent works on hyperbolic space-based recommender systems have shown promising results \cite{HyperML,HVAE,HSCML,HME, HyperSoRc}, particularly with Hyperbolic Graph Convolution Neural Networks (HGCNs) \cite{HGCF,HRCF,HICF}, which combine the advantages of Graph Convolution Neural Networks (GCN) and hyperbolic space.

\begin{figure}[!t]
\centering
\includegraphics[width=1.6in]{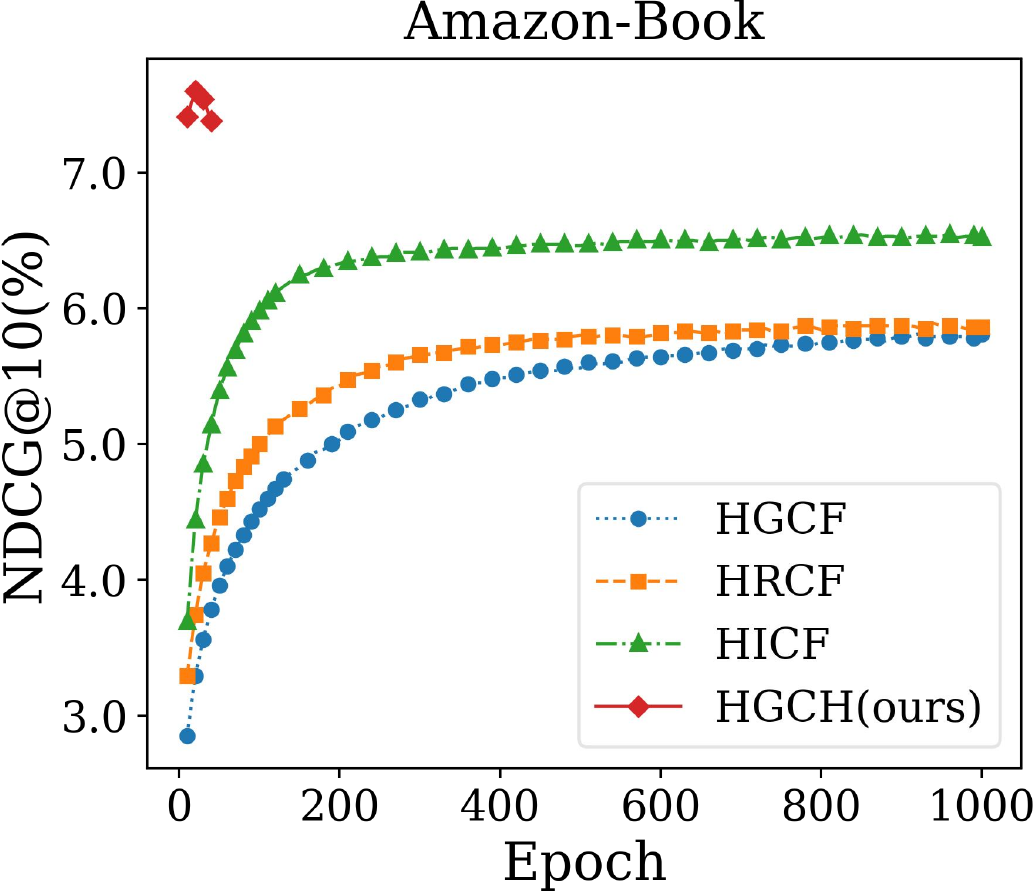}
\includegraphics[width=1.6in]{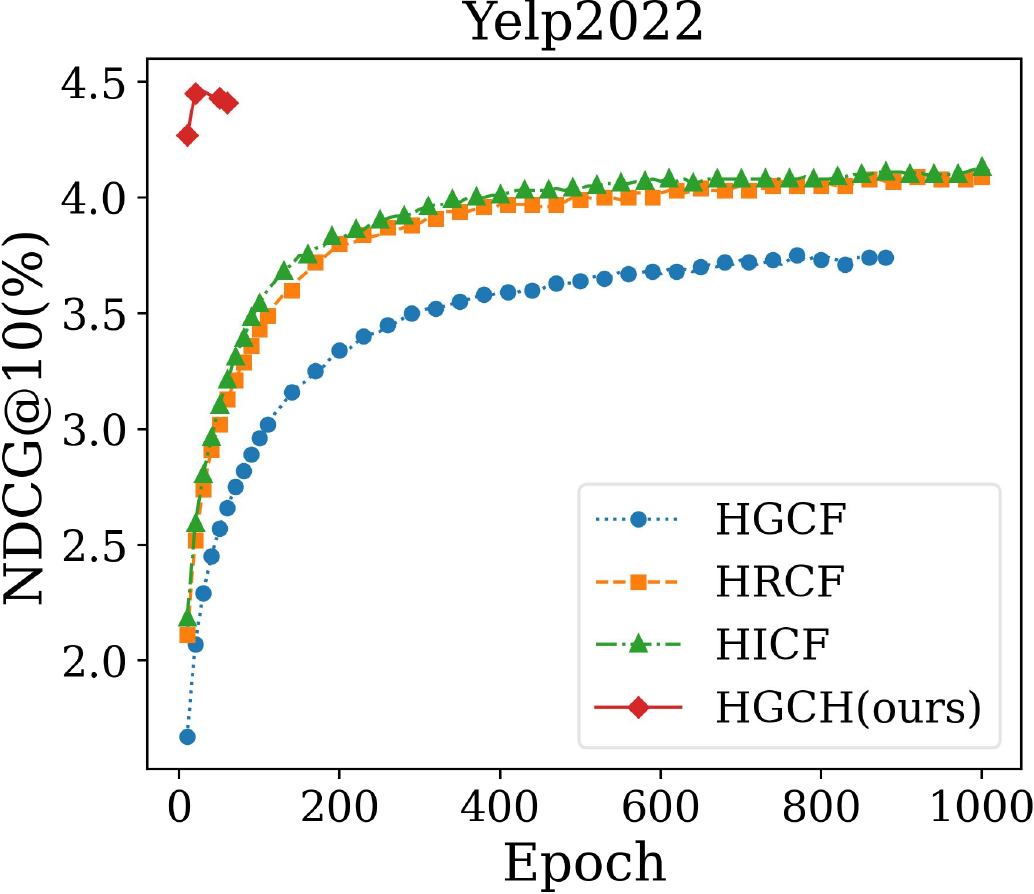}
\caption{NDCG@10 changes with training epochs on validation set of Amazon-Book and Yelp2022. Other metric@$K$s show a similar tendency. HGCH converges very fast so only 40 and 60 epochs were trained on Amazon-Book and Yelp2022, respectively.}
\label{fig:con}
\end{figure}

However, these works suffer from three shortcomings:
\begin{itemize}[leftmargin=*]
\item The designs of these models only partially utilize the nature of hyperbolic space. This is mainly reflected in the following two points: 1) node representations are initialized uniformly \cite{HGCF,HRCF,HICF}, ignoring the different levels of nodes in hyperbolic space; and 2) they utilize the tangent space at the origin to approximate neighbor aggregation, which introduces distortions in relative distances \cite{HGCN}, potentially leading to suboptimal performance.
\item Most works \cite{HGCF,LGCF,HRCF} use a combination of margin ranking loss and uniform random negative sampling, which leads to many invalid triplets in the late training period, resulting in optimization difficulties and slow convergence.
\item They only consider user-item interaction data and neglect other types of side information that can enrich the collaborative graph. Although some works \cite{hu2018leveraging, li2020heterogeneous, shi2020heterogeneous} incorporate side information into a heterogeneous graph, they still fail to capture the intrinsic characteristics of each sub-graph. 
\end{itemize}

To address these challenges, we develop a novel HGCN model for CF with multiple side information, named \textbf{H}yperbolic \textbf{G}raph \textbf{C}ollaborative for \textbf{H}eterogeneous Recommendation (HGCH).
As shown in Fig. \ref{fig:con}, HGCH converges significantly faster and performs better than previous methods. This improvement can be attributed to several important novelties:
1) power law prior-based initialization, which leverages the heterogeneous structure of hyperbolic space and facilitates better training results; 2) hyperbolic neighbor aggregation, which achieves more accurate computation results than tangent space neighbor aggregation; 3) gate fusion with prior, which effectively integrates different side information through prior-based training; 4) and hyperbolic user-specific negative sampling, which facilitates the sampling of a more valid training triplet. Moreover, our model only adds a small number of parameters compared to the state-of-the-art baseline, which further demonstrates the efficiency of our method.
The main contributions of this work are summarized as follows:
\begin{itemize}[leftmargin=*]
\item We propose a new method HGCH, which incorporates power law prior-based initialization, hyperbolic neighbor aggregation and gate fusion with prior to obtain a better distribution of node representations, thus enhancing the utilization of hyperbolic space (\textit{i.e.}, leveraging its exponentially growing capacity to focus more on tail items).
\item We found that the current combination of margin ranking loss and uniform random negative sampling limits the convergence speed of hyperbolic models. To address this, we propose a hyperbolic user-specific negative sampling method, which significantly accelerates the convergence speed of our HGCH compared to other hyperbolic methods. It should be noted that the proposed method is not limited to CF-based models, but is also applicable to other hyperbolic recommendation models.
\item Extensive experiments demonstrate the effectiveness of the proposed method where the maximum recommendation effect on overall items versus all baselines is up to 22.49\%. Our source code is available here: \href{https://github.com/LukeZane118/HGCH}{https://github.com/LukeZane118/HGCH.}
\end{itemize}

\section{Related Work}
\label{sec:relate}
In this section, we review related work from the GCN to HGCN.

\subsection{Graph Convolution Neural Networks}
GCN-based methods have received much attention due to their ability to learn node representations from arbitrary graph structures \cite{gcn,HNLP2}. They are widely used in the fields of computer vision \cite{HCV1,HCV2,HCV3}, natural language processing \cite{HNLP1,HNLP2,HNLP3}, and biocomputing \cite{CB1,CB2,CB3}. There are also some representative GCN approaches in the field of recommender systems. NGCF \cite{NGCF} generalizes GCN to the domain of recommender systems. While LightGCN \cite{LightGCN} further exposes that feature transformations and nonlinear activations in GCN do not play a role in recommendation tasks and thus can be removed. DGCF \cite{DGCF}, on the other hand, attempts to model the user's intent using GCN. 
However, these approaches are mainly built in Euclidean space, which may underestimate the distribution of the user-item dataset. Recent work \cite{HGNN, HGCN, LGCN} has demonstrated the potent power of hyperbolic space to model graph-structured data with hierarchical and power-law distributions.

\subsection{Hyperbolic Recommender Systems}
Hyperbolic recommender systems have been receiving more and more attention lately. HyperML \cite{HyperML} bridges the gap between Euclidean and hyperbolic geometry in recommender systems for the representation of the user and the item through metric learning methods. HAE and HVAE \cite{HVAE} generalize AE and VAE to hyperbolic space for implicit recommendation, respectively. 
HSCML \cite{HSCML} proposes hyperbolic social collaboration metric learning by driving socially relevant users closer to each other. HME \cite{HME} applies hyperbolic space to the point-of-interest recommendation and introduces various side information such as sequential transition, user preference, category, and region information. HyperSoRc \cite{HyperSoRc} learns the representation of users and items in hyperbolic space by introducing social information. 
However, these methods lack the utilization of higher-order connection information of nodes.

Combining graph neural networks, LKGR \cite{LKGR} introduces a know-ledge-aware attention mechanism for hyperbolic recommender systems. HGCF \cite{HGCF} implements multiple layers of skip-connected graph convolutions through tangent spaces to capture high-order connectivity information between nodes. Based on HGCF, HRCF \cite{HRCF} designs a geometry-aware hyperbolic regularizer to facilitate the optimization process. Furthermore, HICF \cite{HICF} proposes an adaptive hyperbolic margin ranking loss. Also based on HGCF, LGCF \cite{LGCF} uses the midpoint of hyperbolic space to realize neighbor aggregation instead of tangent space. However, the designs of these HGCN-based methods do not take full advantage of the nature of hyperbolic space, \textit{i.e.}, simplistic embedding initialization approach and neighbor aggregation by tangent space at the origin.

Furthermore, it should be noted that although LGCF also employs the midpoint in hyperbolic space instead of the tangent space for neighbor aggregation, it involves a transformation process from the Lorentz model to the Klein model \cite{LGCF}. Whether the Einstein midpoint in the Klein model has a geometric interpretation in the Lorentz model remains to be studied \cite{HYBONET}. Moreover, issues regarding the fairness and validity of the experiments conducted in this work have been pointed out.\footnote{Relevant details can be found in the note of \href{https://github.com/marlin-codes/HRCF}{https://github.com/marlin-codes/HRCF.}\label{footnote:note}} In contrast, our approach utilizes the gyromidpoint of the Poincaré model, which allows us to compute the generalized mean of points in hyperbolic space without the need for further conversion to other hyperbolic models. Another similar work is HICF, which proposes a sampling method in hyperbolic space for sampling negative items closer to the positive item, providing additional information for optimization. In contrast, our method samples negative items in hyperbolic space nearer to the user, generating more valid triplets to speed up convergence.

\section{Preliminaries}
\label{sec:preli}
In this section, we first introduce the hyperbolic space approach relevant to this paper and then present our task description. 

\subsection{Hyperbolic Space}
Our goal is to learn low-dimensional embeddings in hyperbolic space, which better capture hierarchical structures than Euclidean space. We use the Poincaré model of hyperbolic geometry, which facilitates computing generalized means of points in hyperbolic space \cite{gyrovector}. In a $d$-dimensional hyperbolic space with curvature $c$ (a negative constant), the negative reciprocal is $k = -1/c$ (non-negative). A hyperbolic space is a Riemannian manifold where the metric depends on the curvature. The tangent space $\mathcal{T}_\mathbf{x}\mathcal{M}$ at a point $\mathbf{x}$ on the manifold $\mathcal{M}$ is a $d$-dimensional Euclidean space approximating the manifold locally at $\mathbf{x}$. The Poincaré model, which is mathematically equivalent to other hyperbolic space representations, is defined by:
\begin{align}
\mathcal{B}^d := \{\mathbf{x} \in \mathbb{R}^d \mid \|\mathbf{x}\|^2 < k\},
\end{align}
where $\|\cdot\|$ denotes the Euclidean norm. The distance in the Poincaré model is given by:
\begin{align}
\label{eq:distance}
d_{\mathcal{B}}(\mathbf{x}, \mathbf{y}) = \sqrt{k} \operatorname{arcosh}\left(1 + 2k \frac{\|\mathbf{x} - \mathbf{y}\|^2}{(k - \|\mathbf{x}\|^2)(k - \|\mathbf{y}\|^2)}\right).
\end{align}

To generalize the operations of Euclidean space to hyperbolic space, we can use the tangent space $\mathcal{T}_\mathbf{x}\mathcal{M}$ as an approximation. The exponential mapping $\text{exp}^k_\mathbf{x}$ and the logarithmic mapping $\text{log}^k_\mathbf{x}$ enable the transformation between the tangent space and the hyperbolic space. We fix the origin $\mathbf{o} = (0, 0, \cdots, 0) \in \mathcal{B}^d$ and use it as a reference point throughout this paper. In the Poincaré method, the exponential mapping $\text{exp}^k_\mathbf{o}$: $\mathcal{T}_\mathbf{o}\mathcal{B}^d \rightarrow \mathcal{B}^d$ is defined as \cite{gyrovector}:
\begin{align}
\label{eq:expmap}
\exp _{\mathbf{o}}^{k}(\mathbf{v})=\tanh (\frac{\|\mathbf{v}\|}{\sqrt{k}}) \frac{\sqrt{k}\mathbf{v}}{\|\mathbf{v}\|},
\end{align}
the logarithmic mapping $\text{log}^k_\mathbf{o}$: $\mathcal{B}^d \rightarrow \mathcal{T}_\mathbf{o}\mathcal{B}^d$ is given by \cite{gyrovector}:
\begin{align}
\label{eq:logmap}
\log _{\mathbf{o}}^{k}(\mathbf{y})=\operatorname{arctanh}(\frac{\|\mathbf{y}\|}{\sqrt{k}}) \frac{\sqrt{k}\mathbf{y}}{\|\mathbf{y}\|}.
\end{align}
\subsection{Task Formulation}
In this section, we present the key concepts and notations used in our work, such as Heterogeneous Collaborative Graph (HCG), high-order node connectivity, and compositional edge relations.

\subsubsection{User-Item Bipartite Graphs}
A common way to represent the historical interactions between users and items (\textit{e.g.}, purchases and visits) in a recommender system is to use a user-item bipartite graph $\mathcal{G}_0=(\mathcal{U} \cup \mathcal{V}, \mathcal{E}_0)$, where $\mathcal{U}$ and $\mathcal{V}$ denote the user set and item set, respectively, and an edge $y_{u,v} = 1$ in edge set $\mathcal{E}_0$ indicates that user $u$ has interacted with item $v$; otherwise $y_{u,v} = 0$.

\begin{figure}[!t]
    \centering
    \includegraphics[width=8cm]{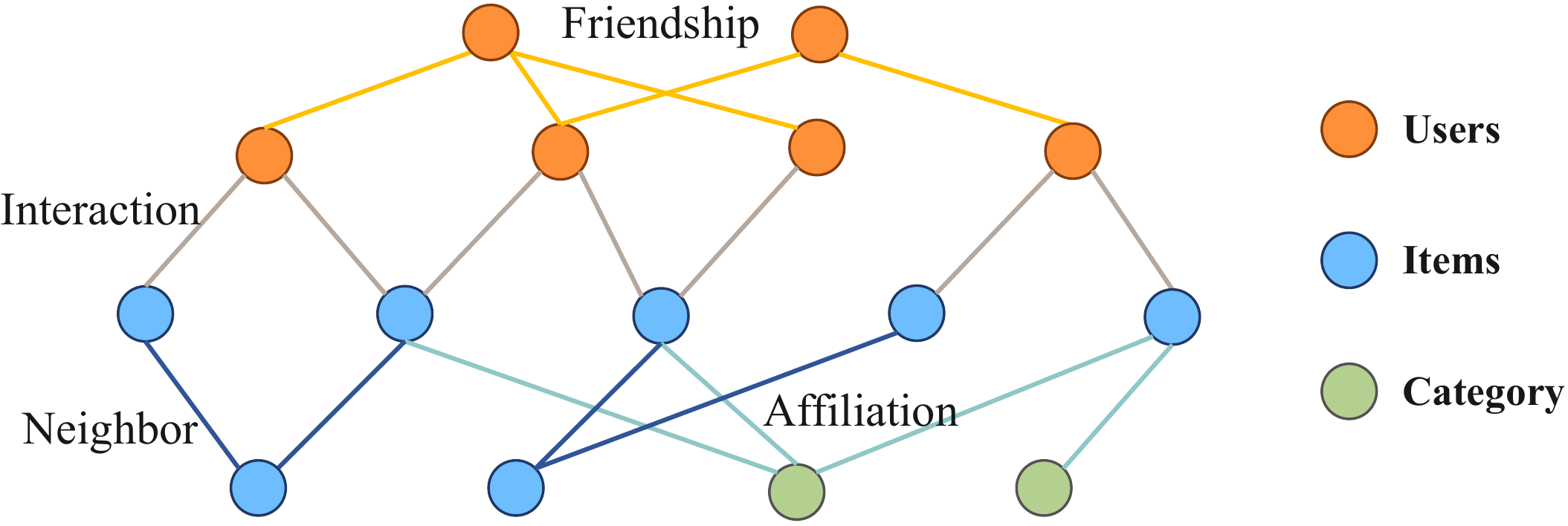}
    \caption{An Example of a Heterogeneous Collaborative Graph. Its node set contains the user, item, and category; its edge set contains the user-item link, user-user link, item-item link and item-category link.}
    \label{fig:ccg}
\end{figure}
    
\subsubsection{Heterogeneous Collaborative Graph}
We define a heterogeneous collaborative graph as $\mathcal{G}=\mathcal{G}_0 \cup \mathcal{G}'=({\mathcal{H}, \mathcal{E}})$, where $\mathcal{H}$ is the node set that includes user nodes $\mathcal{U}$, item nodes $\mathcal{V}$ and other optional nodes such as category nodes etc. $\mathcal{E}$ is the edge set that includes user-item edges $\mathcal{E}_{0}$ and other types of edges,  and $\mathcal{G}'=\{\mathcal{G}_1, \mathcal{G}_2, \cdots, \mathcal{G}_{|\mathcal{S}|-1} \}$ represents the set of all side-information graphs. A heterogeneous collaborative graph $\mathcal{G}$ is also associated with a node type mapping function $\phi: \mathcal{H} \rightarrow \mathcal{T}$ and an edge type mapping function $\psi: \mathcal{E} \rightarrow \mathcal{S}$. $\mathcal{T}$ and $\mathcal{S}$ denote predefined node types and edge types, respectively, where $|\mathcal{T}|+|\mathcal{S}|>2$. Fig. \ref{fig:ccg} shows an example of the HCG.

\subsubsection{Problem Statement} 
Based on the above definitions, we can formulate the recommendation problem that we aim to address in this paper as follows:
\begin{itemize}[leftmargin=*]
\item \textbf{Input}: a heterogeneous collaborative graph $\mathcal{G}$ comprising a user-item interaction graph $\mathcal{G}_0$ and a side information subgraph $\mathcal{G}'$.
\item \textbf{Output}: a function that estimates the probability $\hat{y}_{u,v}$ of user $u$ accepting item $v$.
\end{itemize}

\section{Method}
\label{sec:method}
\begin{figure*}
    \centering
    \includegraphics[width=17cm]{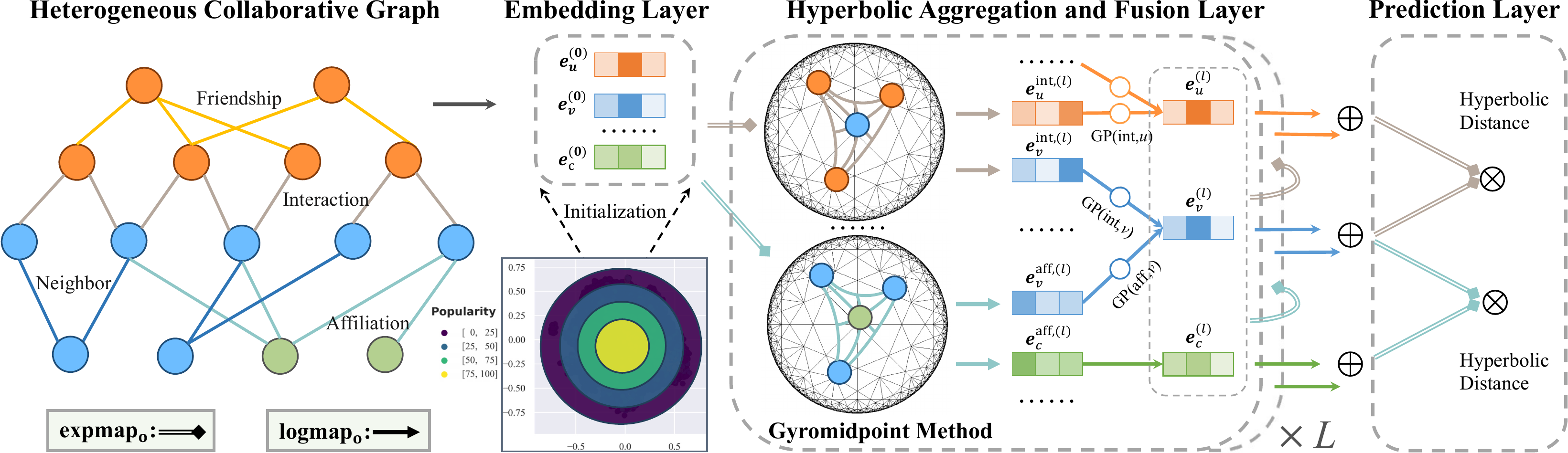}
    \caption{An illustration of HGCH model architecture. It consists of three main components: 1) power law prior-based embedding layer, 2) hyperbolic aggregation and fusion layer, and 3) prediction layer. Note that "int" and "aff" denote the subspace "interaction" and "affiliation", respectively. "GP" means gate fusion with prior.}
    \label{fig:framework}
\end{figure*}

We now propose the HGCH model, which leverages hyperbolic geometry to capture complex and hierarchical relationships in recommender systems. The model consists of three main components, as illustrated in Fig. \ref{fig:framework}: 
1) a power law prior-based embedding layer initializing nodes in tangent space by popularity; 2) a hyperbolic aggregation and fusion layer updating node embeddings through prior-based gating; and 3) a prediction layer that computes matching scores between nodes in the hyperbolic space. In addition, our HGCH also includes a hyperbolic user-specific negative sampling for accelerating the training convergence speed.

\subsection{Embedding Layer with Power Law Prior}
Most existing hyperbolic recommender systems define node embeddings in the hyperbolic space and train them with a Riemannian optimizer \cite{HME, HGCF, HICF, HRCF}. However, as reported by AMCAD \cite{AMCAD} and confirmed by our initial experiments, this approach does not offer significant advantages over defining and training node embeddings in the tangent space (Euclidean space) with a Euclidean optimizer. Therefore, we adopt the latter approach in our model.
The initialization of node embeddings is crucial for the optimization process and the final performance \cite{Xavier}. A good initialization should reflect the inherent properties of the data, while a bad initialization may lead to suboptimal solutions. Previous works \cite{HME, HGCF, HICF, HRCF} initialized node embeddings by sampling from uniform or Gaussian distributions in the tangent space. However, this ignores the fact that node popularity is correlated with the norm of the hyperbolic embedding, such that more popular nodes tend to have smaller norms and are closer to the origin \cite{HME, HGCF}. A uniform or Gaussian prior distribution for all nodes probably results in misplacing popular nodes far from the origin, which may hamper the model's ability to learn meaningful representations.

To address this issue, we propose to initialize node embeddings based on a power law prior distribution, which assigns different initialization ranges to nodes according to their popularity. In this way, nodes with higher popularity are more likely to be initialized near the origin, while nodes with lower popularity are more likely to be initialized away from the origin. Formally, given a uniform distribution, the power law prior-based sampling distribution of node $n$ is:
\begin{align}
\mathbf{e}^{(0)}_n \sim \text{Uni}(- a x_n^{-b}, a x_n^{-b}),
\end{align}
where $x_n$ denotes the frequency of node $n$, $a$ is the scale of initialization consistent with uniform initialization, and $b$ is the hyperparameter of power law distribution. Note that since $\text{exp}^k_\mathbf{o}$ and $\text{log}^k_\mathbf{o}$ are both conformal \cite{gyrovector}, the prior distribution is preserved when mapping embeddings between tangent and hyperbolic spaces.

\subsection{Hyperbolic Aggregation and Fusion Layers}
Next, we establish a novel hyperbolic graph convolution layer that recursively propagates and fuses embeddings from different subspaces through high-order connectivity. In this section, we first introduce the two components of our layer: hyperbolic neighbor aggregation and gate fusion with prior. Then we explain how to stack multiple layers for deeper representation learning.
\subsubsection{Hyperbolic Neighbor Aggregation}
Previous work \cite{HGCN} extends the Euclidean GCN neighbor aggregation to hyperbolic space by the tangent space method. It consists of three steps: 1) project the hyperbolic space embedding to the tangent space at the origin by $\text{log}^k_\mathbf{o}$; 2) apply the Euclidean GCN neighbor aggregation in the tangent space; and 3) project the embedding back to the hyperbolic space by $\text{exp}^k_\mathbf{o}$. In this way, the GCN neighbor aggregation in hyperbolic space can be approximated. Formally, given a hyperbolic space embedding of node $n$, its tangent space neighbor aggregation proceeds according to the following equations:
\begin{align}
&\mathbf{e}^{(l-1)}_n=\log _\mathbf{o}^{k}(\mathbf{h}^{(l-1)}_n),\\
&\mathbf{e}^{(l)}_n= \sum\limits_{i \in \mathcal{N}_n}  \frac{1}{|\mathcal{N}_n|} \mathbf{e}^{(l-1)}_i,\\
&\mathbf{h}^{(l)}_n=\exp _\mathbf{o}^{k}(\mathbf{e}^{(l)}_n),
\end{align}
where $\mathcal{N}_n$ denotes the set containing node $n$ itself and its neighbors. However, this approach has such a drawback: it performs neighbor aggregation of all nodes in the same tangent space (the origin), instead of their own tangent spaces, which introduces distortions in relative distance \cite{HGCN}, thus may lead to a suboptimal recommendation effect. 
To address this issue, we propose to directly implement neighbor aggregation in hyperbolic space using the gyromidpoint method \cite{gyrovector}, which computes the weighted average of a set of points in hyperbolic space as follows:
\begin{align}
\label{eq:midpoint}
m\left(\mathbf{x}_{1}, \cdots, \mathbf{x}_{N} ; \boldsymbol{\nu} \right)=\frac{1}{2} \otimes_{k}\left[\sum_{i=1}^{N} \frac{\nu_{i} \lambda(k,{\mathbf{x}_i})}{\sum_{j=1}^{N} \nu_{j}\left(\lambda(k,{\mathbf{x}_j})-1\right)} \mathbf{x}_{i}\right],
\end{align}
where $\nu_i$ is the weight of node $\mathbf{x}_i$, $\lambda(k,{\mathbf{x}})=2\left(1-\frac{\|\mathbf{x}\|^{2}}{k}\right)^{-1}$ ($\|\cdot\|$ is Euclidean norm), and $\otimes_k$ is the Möbius scalar multiplication \cite{gyrovector}. For simplicity, we set $\nu_i$ to 1 for all nodes, meaning that they have equal importance. Other methods, such as attention mechanism \cite{ attention}, can be used to learn different weights for different nodes, but this is beyond the scope of this paper.
The gyromidpoint method is a natural generalization of the Euclidean midpoint to hyperbolic space. It preserves the distance between points and their midpoint, and it is invariant under isometries of hyperbolic space. So it can be seen as a way of performing neighbor aggregation in hyperbolic space by averaging the information from neighboring nodes.

Hence, given the tangent space embedding of a node $n$ of $(l-1)$-th layer, we want to get the neighbor information of $n$ from subgraph $\mathcal{G}'$. We first project it into the hyperbolic subspace $s$ by $\text{exp}^k_\mathbf{o}$:

\begin{align}
\label{eq:expmap4mchg}
\mathbf{h}^{s,(l-1)}_n=\exp _\mathbf{o}^{k_s}(\mathbf{e}^{(l-1)}_n).
\end{align}
The exponential map $\text{exp}^k_\mathbf{o}$ maps a point in the tangent space at the origin $\mathbf{o}$ to a point in hyperbolic space with curvature $k$. It preserves the distance and angle between vectors in the tangent space and their images in hyperbolic space. It can be seen as a way of projecting information from Euclidean space to hyperbolic space by preserving the geometry. Then we perform neighbor aggregation by Eq. \eqref{eq:midpoint}:
\begin{align}
\mathbf{h}^{s,(l)}_n=\frac{1}{2} \otimes_{k_s}\left[\sum_{i \in \mathcal{N}_n} \frac{ \lambda({k_s},{\mathbf{h}^{s,(l-1)}_i})}{\sum_{j \in \mathcal{N}_n} \left(\lambda({k_s},{\mathbf{h}^{s,(l-1)}_j})-1\right)} \mathbf{h}^{s,(l-1)}_n \right],
\end{align}
thus aggregating information from neighboring nodes directly in hyperbolic space. 

\subsubsection{Multi-Space Information Fusion}
\label{sec:msif}
After neighbor aggregation in each subspace, we need to fuse the output from each subspace to update the embedding of the nodes. Since the curvature of each subspace is not necessarily the same, we need to project the embedding to a uniform tangent space first, as follows:

\begin{align}
\mathbf{e}^{s,(l)}_n=\log _\mathbf{o}^{k_s}(\mathbf{h}_n^{s,(l)}).
\end{align}
Then we can consider the fusion of multiple information in the tangent space, which is:
\begin{align}
\mathbf{e}^{(l)}_n=\sum_{s \in \mathcal{S}_n}\mathbf{g}_n^s \odot \mathbf{e}^{s,(l)}_n,
\end{align}
where $\mathcal{S}_n$ is the set of subspaces that node $n$ involves in, $\odot$ is element-wise product and $\mathbf{g}_n^s$ is the weight of embedding $\mathbf{e}_n^s$. An intuitive idea is to set a learnable weight to control the weight of each kind of information automatically. Accordingly, we use the gate mechanism to implement this idea, which we call gate fusion. It is formulated as follows:
\begin{align}
\mathbf{g}_{\text{gate}}=\frac{\sigma(\mathbf{W}_{t_n}^{s}\mathbf{e}_n^{(0)}/\|\mathbf{e}_n^{(0)}\|)}{\sum_{i \in \mathcal{S}_n}\sigma(\mathbf{W}_{t_n}^i\mathbf{e}_n^{(0)}/\|\mathbf{e}_n^{(0)}\|)},
\end{align}
where $t_n$ is the type of node $n$, $\textit{i.e.}$, $t_n \in \{\textit{user}, \textit{item}, \textit{category}, \cdots \} $, $\mathbf{W}^s_{t_n}$ is the learnable parameter of the gate of subspace $s$ corresponding to the type of node $n$, and $\sigma$ is the sigmoid function. The gate mechanism is a way of learning adaptive weights for different kinds of information based on their relevance to the node type and their initial embedding. The sigmoid function ensures that the weights are between 0 and 1, and normalized by their sum. The gate mechanism can help select useful information and filter out noisy information for each node type. However, in practice, we found that gate fusion can easily fall into the local optimum and even perform less well than the model without side information in some cases. So we directly use normalized neighbor numbers of node $n$ in subspace $s$ as a prior weight, namely prior-based fusion, and it is given by:
\begin{align}
\mathbf{g}_{\text{prior}} = \frac{ |{\mathcal{N}_n^s}|}{\sum_{i \in {\mathcal{S}_n}} |{\mathcal{N}_n^i}|}.
\end{align}
The prior-based fusion is a way of assigning weights for different kinds of information based on their frequency or density in each subspace. The intuition is that more neighbors imply more information and thus higher importance. The prior-based fusion can help balance different kinds of information and avoid overfitting or underfitting problems. What is more, if the gate fusion can learn based on prior instead of learning from scratch, the performance may get further improvement. Therefore, we combine the two and call this approach the gate fusion with prior, which is defined as:
\begin{align}
\mathbf{g}_{\text{gate\&prior}}=\frac{ |{\mathcal{N}_n^s}|\sigma(\mathbf{W}_{t_n}^s\mathbf{e}_n^{(0)}/\|\mathbf{e}_n^{(0)}\|)}{\sum_{i \in \mathcal{S}_n} |{\mathcal{N}_i^s}|\sigma(\mathbf{W}_{t_n}^i\mathbf{e}_n^{(0)}/\|\mathbf{e}_n^{(0)}\|)}.
\end{align}
According to experiments, gate fusion with prior is better and stable in most cases, so we adopt this structure for HGCH. The empirical comparison of these structures is shown in Section \ref{sec:cofm}. 
Although we avoid the distortion of neighbor aggregation, we still inevitably bring distortion when introducing side information since it is difficult to directly aggregate nodes in different hyperbolic spaces with different curvatures. However, it is confirmed by Section \ref{sec:cofm} that the benefits of introducing side information far outweigh the drawback. Besides, we also use the learnable gate method as a remedy for the distortion. If both neighbor aggregation and information fusion are implemented in tangent space, two distortions will be accumulated.
\subsubsection{High-Ordrer Propagation}
Further, we can stack multiple layers to explore high-order connectivity information and collect information from multi-hop neighbors. 
However, previous work \cite{HGCF} found that stacking multiple layers leads to degradation in the model's overall performance due to gradient disappearance and over smoothing. Here, we refer to HGCF \cite{HGCF} and use SkipGCN to aggregate the output of multiple layers, which is defined as follows:
\begin{align}
\mathbf{e}_{n}=\sum_{i=1}^{L}\mathbf{e}_{n}^{(i)},
\end{align}
where $L$ is the number of convolutional layers. The SkipGCN is a way of combining information from all layers by adding skip connections from each layer to the final layer. The skip connections can help preserve the original information and avoid the loss of gradient and information during the propagation process. In this way, the final output of the convolutional layer embedding contains rich neighbor information and side information, and we next employ the hyperbolic distance $d_\mathcal{B}$ to evaluate the similarity of node pairs.

\subsection{Prediction Layer}
The embeddings of the final output are first projected into the hyperbolic space of the scoring layer, and then we use the hyperbolic distance $d_\mathcal{B}$ defined in Eq. \eqref{eq:distance} to measure the similarity of the projected embeddings $\mathbf{h}_i$ and $\mathbf{h}_j$, as follows:
\begin{align}
&\mathbf{h}_i=\exp _\mathbf{o}^{k_{s}}(\mathbf{e}_i),\\
&\mathbf{h}_j=\exp _\mathbf{o}^{k_{s}}(\mathbf{e}_j),\\
&\hat{y}_{i, j}=-d_{\mathcal{B}}(\mathbf{h}_i, \mathbf{h}_j)^2.
\end{align}
The hyperbolic distance $d_{\mathcal{B}}$ is based on the inner product of two points in hyperbolic space. It reflects the similarity between two nodes based on their hyperbolic embeddings. This completes the prediction of the node pair ratings.

\subsection{Optimization}
We adopt the margin ranking loss used by HGCF \cite{HGCF} to optimize our model. The margin ranking loss for the target subgraph $\mathcal{G}_0$ is given by:
\begin{align}
\mathcal{L}_\text{CF}=\mathcal{L}_{\mathcal{G}_0}=\max \left(\hat{y}_{u,j}-\hat{y}_{u,i}+m, 0\right),
\end{align}
where $m$ is a margin parameter, item $i$ is a positive sample drawn from $y_{u,i}=1 \in \mathcal{G}_0$, and item $j$ is a negative sample drawn from $\mathcal{V}$. The margin $m$ specifies the minimum difference required between positive and negative sample scores. As $m$ approaches infinity, the margin ranking loss becomes equivalent to the BPR Loss \cite{BPR}. However, this causes the model to focus too much on easy pairs, resulting in slow training and poor performance due to many embeddings clustering near the boundaries of the ball space.
For other subgraphs $\mathcal{G}'$, we also train their embeddings by margin ranking loss:

\begin{align}
\mathcal{L}_\text{SI}=\sum_{\mathcal{G}_{i} \in \mathcal{G}'} \mathcal{L}_{\mathcal{G}_{i}},
\end{align}
where node $\mathcal{G}_{i}$ denotes subgraphs of each link type, and we only sample negative nodes in subgraph $\mathcal{G}_{i}$ when calculating $\mathcal{L}_{\mathcal{G}_{i}}$.  Eventually, our objective function is a multitask loss with $\mathcal{L}_\text{CF}$ as the main task and $\mathcal{L}_\text{SI}$ as the auxiliary task:

\begin{align}
\mathcal{L}=\mathcal{L}_\text{CF} + \alpha \mathcal{L}_\text{SI},
\end{align}
where $\alpha$ is the hyperparameter that controls the weight of the auxiliary task $\mathcal{L}_\text{SI}$. 

\begin{algorithm}[t]
\caption{Hyperbolic User-specific Negative Sampling}
\label{alg:sampling}
\begin{flushleft}
{\bf Input:}
Hyperparameters $n_\text{neg}$; The item set $\mathcal{V}$; The embedding matrix $\mathbf{E}$; The index of the user $u$ and his/her positive item $\mathcal{V}_u$ in the training set.\\
{\bf Output:} 
The item index of the negative sample.
\end{flushleft}
\begin{algorithmic}[1]
\STATE $\mathcal{V}^\text{neg}_u \gets$ randomly sample $n_\text{neg}$ items from $\mathcal{V} \backslash \mathcal{V}_u$
\STATE $\textit{minDist} \gets \infty$
\STATE $\textit{minIndex} \gets -1$
\STATE Get the tangent embeddings of $u$ the from $\mathbf{E}$, \textit{i.e.} $\mathbf{e}_u$
\STATE $\mathbf{h}_u \gets \exp _\mathbf{o}^{k}(\mathbf{e}_u)$ 
\FOR{$i \in \mathcal{V}^\text{neg}_u$}
    \STATE Get the tangent embeddings of $i$ the from $\mathbf{E}$, \textit{i.e.} $\mathbf{e}_i$
    \STATE $\mathbf{h}_i \gets \exp _\mathbf{o}^{k}(\mathbf{e}_i)$
    \STATE $d_{u,i} \gets d_{\mathcal{B}}(\mathbf{h}_u, \mathbf{h}_i)$ 
    \IF{$d_{u,i} < \textit{minDist}$}
        \STATE $\textit{minDist} \gets d_{u,i}$
        \STATE $\textit{minIndex} \gets i$
    \ENDIF
\ENDFOR
\STATE \textbf{return} $\textit{minIndex}$
\end{algorithmic}
\end{algorithm}

Moreover, in our preliminary experiments, we observed that HGCN-based methods tend to converge slowly. This phenomenon is likely due to the inefficiency of random sampling techniques for sampling training triplets in this setting. It is easy to see that, during the later stages of training, most negative pairs ($u$, $j$) and positive pairs ($u$, $i$) already have a distance larger than the margin $m$, making them ineffective for optimization. Furthermore, random sampling assumes a uniform distribution, which treats all items equally, leading to most sampled training triplets being invalid for model optimization. To address this issue, we need a user-specific, self-optimizing, and data-independent method. Alg. \ref{alg:sampling} shows our approach. In each iteration, we randomly pick $n_\text{neg}$ items $\mathcal{V}^\text{neg}_u$ and compute their hyperbolic distances to user $u$. Then, we keep the item with the smallest distance, which means it is close to user $u$ in the hyperbolic space. This strategy guarantees that the sampled negative node remains close to user $u$, thereby facilitating the sampling of valid training triplets.

\section{Discussion}
\label{sec:discus}
\subsection{Space Complexity Analysis}
HGCH's space complexity has two main parts: 1) node embedding, $\mathcal{O}(|\mathcal{H}|d)$; and 2) gate, $\mathcal{O}(|\mathcal{G}|d^2)$. Hence the total space complexity is $\mathcal{O}((|\mathcal{H}| + |\mathcal{G}|d)d)$. When only using user-item interaction data, the space complexity of HGCH is $\mathcal{O}(|\mathcal{H}|d)$, which is consistent with the lightweight models having only embedding parameters such as LightGCN \cite{LightGCN} and HGCF \cite{HGCF}; when including side information, the numbers of new node from side graph are generally small in real scenarios. Besides, it is often $|\mathcal{H}| \gg |\mathcal{G}|d$, so HGCH only adds a small number of parameters in this case.

\subsection{Relation to Heterogeneous Graph Methods}
Heterogeneous graph recommendations often use path-based methods, which extract high-order path information for the model. They either select salient paths \cite{PS1,PS2} or use predefined meta-paths \cite{MP1,MP2,sun2011pathsim,jin2020efficient} to handle multiple node relationships. However, the paths are not optimized for the recommendation goals, or they need domain knowledge and may not be optimal. Traditional methods struggle with large and complex nodes and relations \cite{KGAT}. Our approach separately processes and fuses different node relationships by hyperbolic subspace and multi-space information fusion, and then combines multiple relationship paths by multi-layer GCN's high-order connectivity.

\begin{table*}[htbp]
\footnotesize
  \centering
  \caption{Performances for all datasets. The best performing model on each dataset and metric is highlighted in bold, and second best model is underlined. Asterisks denote statistically significant wilcoxon signed rank test for the difference in scores between the best and second-best models.}
  \setlength{\tabcolsep}{1.4mm}{
    \begin{tabular}{cc|ccccc|cccc|cccc|cc}
    \toprule
    Datasets & Metric & BPRMF & WRMF  & NGCF  & LightGCN & DGCF  & HGCF  & HRCF  & HICF  & HGCH  & HGCF+ & HRCF+ & HICF+ & HGCH+ & $\Delta$(\%) & $\Delta_+$(\%) \\
    \midrule
    \multirow{4}[1]{*}{Amazon-CD} & R@10  & 0.0666  & 0.0762  & 0.0710  & 0.0834  & 0.0765  & 0.0975  & 0.1015  & \underline{0.1031} & \textbf{0.1067*} & 0.1055  & 0.0983  & 0.1009  & \textbf{0.1196*} & 3.49\% & 16.00\% \\
          & R@20  & 0.1036  & 0.1147  & 0.1092  & 0.1259  & 0.1138  & 0.1436  & 0.1476  & \underline{0.1522} & \textbf{0.1555*} & 0.1585  & 0.1483  & 0.1518  & \textbf{0.1743*} & 2.17\% & 14.52\% \\
          & N@10  & 0.0553  & 0.0631  & 0.0586  & 0.0688  & 0.0633  & 0.0823  & 0.0843  & \underline{0.0867} & \textbf{0.0904*} & 0.0872  & 0.0798  & 0.0824  & \textbf{0.1010*} & 4.27\% & 16.49\% \\
          & N@20  & 0.0672  & 0.0756  & 0.0710  & 0.0825  & 0.0754  & 0.0969  & 0.0989  & \underline{0.1021} & \textbf{0.1057*} & 0.1042  & 0.0960  & 0.0988  & \textbf{0.1182*} & 3.53\% & 15.77\% \\
    \midrule
    \multirow{4}[1]{*}{Amazon-Book} & R@10  & 0.0666  & 0.0609  & 0.0637  & 0.0827  & 0.0714  & 0.0983  & 0.0996  & \underline{0.1099} & \textbf{0.1261*} & 0.1008  & 0.0996  & 0.1115  & \textbf{0.1320*} & 14.74\% & 20.11\% \\
          & R@20  & 0.0998  & 0.0927  & 0.0974  & 0.1215  & 0.1069  & 0.1413  & 0.1434  & \underline{0.1553} & \textbf{0.1701*} & 0.1451  & 0.1434  & 0.1577  & \textbf{0.1797*} & 9.53\% & 15.71\% \\
          & N@10  & 0.0537  & 0.0511  & 0.0523  & 0.0690  & 0.0591  & 0.0835  & 0.0843  & \underline{0.0947} & \textbf{0.1118*} & 0.0845  & 0.0833  & 0.0953  & \textbf{0.1160*} & 18.06\% & 22.49\% \\
          & N@20  & 0.0645  & 0.0614  & 0.0632  & 0.0815  & 0.0706  & 0.0972  & 0.0984  & \underline{0.1090} & \textbf{0.1251*} & 0.0987  & 0.0973  & 0.1099  & \textbf{0.1305*} & 14.77\% & 19.72\% \\
    \midrule
    \multirow{4}[2]{*}{Gowalla} & R@10  & 0.1075  & 0.0999  & 0.1143  & 0.1286  & 0.1204  & 0.1296  & 0.1303  & \underline{0.1342} & \textbf{0.1447*} & 0.1392  & 0.1409  & 0.1453  & \textbf{0.1576*} & 7.82\% & 17.44\% \\
          & R@20  & 0.1578  & 0.1499  & 0.1697  & 0.1853  & 0.1755  & 0.1912  & 0.1914  & \underline{0.1959} & \textbf{0.2086*} & 0.2053  & 0.2061  & 0.2114  & \textbf{0.2273*} & 6.48\% & 16.03\% \\
          & N@10  & 0.1035  & 0.0926  & 0.1110  & 0.1262  & 0.1171  & 0.1237  & 0.1247  & \underline{0.1289} & \textbf{0.1396*} & 0.1343  & 0.1351  & 0.1402  & \textbf{0.1532*} & 8.30\% & 18.85\% \\
          & N@20  & 0.1188  & 0.1093  & 0.1277  & 0.1432  & 0.1338  & 0.1420  & 0.1428  & \underline{0.1471} & \textbf{0.1581*} & 0.1538  & 0.1546  & 0.1598  & \textbf{0.1734*} & 7.48\% & 17.88\% \\
    \midrule
    \multirow{4}[2]{*}{Yelp2022} & R@10  & 0.0500  & 0.0629  & 0.0546  & 0.0655  & 0.0609  & 0.0667  & 0.0718  & \underline{0.0733} & \textbf{0.0790*} & 0.0666  & 0.0657  & 0.0679  & \textbf{0.0795*} & 7.78\% & 8.46\% \\
          & R@20  & 0.0838  & 0.1017  & 0.0897  & 0.1057  & 0.0986  & 0.1109  & 0.1191  & \underline{0.1206} & \textbf{0.1273*} & 0.1108  & 0.1099  & 0.1128  & \textbf{0.1280*} & 5.56\% & 6.14\% \\
          & N@10  & 0.0409  & 0.0514  & 0.0448  & 0.0552  & 0.0508  & 0.0549  & 0.0595  & \underline{0.0606} & \textbf{0.0660*} & 0.0549  & 0.0538  & 0.0557  & \textbf{0.0670*} & 8.91\% & 10.56\% \\
          & N@20  & 0.0520  & 0.0642  & 0.0563  & 0.0682  & 0.0629  & 0.0693  & 0.0748  & \underline{0.0758} & \textbf{0.0814*} & 0.0692  & 0.0681  & 0.0702  & \textbf{0.0824*} & 7.39\% & 8.71\% \\
    \bottomrule
    \end{tabular}%
    }
  \label{tab:main}%
\end{table*}%

\section{Experiments}
\label{sec:experi}
We evaluate our proposed approach on four real data sets. We want to answer the following research questions:
\begin{itemize}[leftmargin=*]
\item \textbf{RQ1}: How does HGCH perform compared to the state-of-the-art recommender methods?
\item \textbf{RQ2}: What is the impact of each component?
\item \textbf{RQ3}: How well do different fusion methods work on HGCH?
\item \textbf{RQ4}: What kind of embedding has HGCH method learned?
\item \textbf{RQ5}: How about the convergent speed of HGCH?
\end{itemize}

\subsection{Dataset Description}
We evaluate the proposed method on four public available datasets: Amazon-CD\footnote{\href{http://jmcauley.ucsd.edu/data/amazon}{http://jmcauley.ucsd.edu/data/amazon.}\label{url:amazon}}, Amazon-Book\textsuperscript{\ref{url:amazon}}, Gowalla\footnote{\href{http://snap.stanford.edu/data/loc-gowalla.html}{http://snap.stanford.edu/data/loc-gowalla.html.}}, and Yelp2022\footnote{\href{https://www.yelp.com/dataset}{https://www.yelp.com/dataset.}}, each varying in domain, size, sparsity, and side information. We had to processe the raw data from scratch to include side information, as previous studies \cite{NGCF, LightGCN} only offered processed interaction datasets. 

\subsection{Experimental Settings}
\subsubsection{Evaluation Metrics}
We adopt two widely-used evaluation metrics Recall@$K$ and NDCG@$K$ \cite{jarvelin2002cumulated}, with $K$ set to 10 and 20. Our experiments are conducted on the RecBole \cite{recbole1.0} framework, following its implementation of Recall@$K$ and NDCG@$K$.

\subsubsection{Baselines}
To show the effectiveness of our model, we compare our proposed HGCH with latent factor-based (BPRMF \cite{BPR} and WRMF \cite{WRMF}), GCN-based (NGCF \cite{NGCF}, LightGCN \cite{LightGCN}, and DGCF \cite{DGCF}), and HGCN-based (HGCF \cite{HGCF}, HRCF \cite{HRCF}, and HICF \cite{HICF}). Besides, we refer to the version using HCG directly as HGCF+, HRCF+, and HICF+, respectively. 
For our model, we denote the version that uses only the user-item bipartite graph and the version that utilizes HCG as HGCH and HGCH+, respectively. Note that the main difference between HGCF+/HRCF+/HICF+ and HGCH+ is the neighbor aggregation, the former proceeds directly on the HCG, but the latter performs on each subgraph of the HCG separately.

We do not include LGCF \cite{LGCF} as it has been noted to have issues\textsuperscript{\ref{footnote:note}}. In addition, although there are more recent GCN-based methods \cite{ultragcn,svdgcn}, we primarily focus on LightGCN, as it serves as the backbone for HGCF, HRCF, HICF, and our method. 
These newer GCN-based methods can be integrated with ours.

\subsubsection{Hyperparameter Settings}
We followed the original settings and tuned the parameters for each model. 
Compared to previous work that generally trained for 500 epochs \cite{HGCF,HRCF,HICF}, we extended the training to 1000 epochs because we found many methods had not yet converged. Additionally, consistent with prior work, we employed an early stopping strategy, meaning that training was stopped early if the NDCG@10 on the validation set did not improve for 100 consecutive epochs.

\begin{table*}[!t]
\footnotesize
  \centering
  \caption{Evaluation of HGCH with different subsets of components for all datasets.}
  \setlength{\tabcolsep}{1.9mm}{
    \begin{tabular}{cccc|cccc|cccc}
    \toprule
    \multicolumn{4}{c|}{Component} & \multicolumn{4}{c|}{Amazon-CD} & \multicolumn{4}{c}{Amazon-Book} \\
    \midrule
    HS    & PI    & HA    & GP    & Recall@10  & Recall@20  & NDCG@10  & NDCG@20  & Recall@10  & Recall@20  & NDCG@10  & NDCG@20 \\
    \midrule
          &       &       &       & 0.0930 (0.0\%) & 0.1375 (0.0\%) & 0.0785 (0.0\%) & 0.0926 (0.0\%) & 0.0921 (0.0\%) & 0.1304 (0.0\%) & 0.0797 (0.0\%) & 0.0919 (0.0\%) \\
    \checkmark     &       &       &       & 0.0960 (3.2\%) & 0.1390 (1.1\%) & 0.0827 (5.4\%) & 0.0960 (3.7\%) & 0.1062 (15.3\%) & 0.1431 (9.7\%) & 0.0950 (19.2\%) & 0.1062 (15.6\%) \\
          & \checkmark     & \checkmark     &       & 0.0984 (5.8\%) & 0.1451 (5.5\%) & 0.0830 (5.7\%) & 0.0976 (5.4\%) & 0.1025 (11.3\%) & 0.1447 (11.0\%) & 0.0881 (10.5\%) & 0.1015 (10.4\%) \\
    \checkmark     &       & \checkmark     &       & 0.1052 (13.1\%) & 0.1528 (11.1\%) & 0.0891 (13.5\%) & 0.1039 (12.2\%) & 0.1230 (33.6\%) & 0.1671 (28.1\%) & 0.1092 (37.0\%) & 0.1226 (33.4\%) \\
    \checkmark     & \checkmark     &       &       & 0.1071 (15.2\%) & 0.1569 (14.1\%) & 0.0908 (15.7\%) & 0.1064 (14.9\%) & 0.1253 (36.0\%) & 0.1704 (30.7\%) & 0.1111 (39.4\%) & 0.1248 (35.8\%) \\
    \checkmark     & \checkmark     & \checkmark     &       & 0.1067 (14.7\%) & 0.1555 (13.1\%) & 0.0904 (15.2\%) & 0.1057 (14.1\%) & 0.1261 (36.9\%) & 0.1701 (30.4\%) & 0.1118 (40.3\%) & 0.1251 (36.1\%) \\
    \checkmark     & \checkmark     & \checkmark     & \checkmark     & 0.1196 (28.6\%) & 0.1743 (26.8\%) & 0.1010 (28.7\%) & 0.1182 (27.6\%) & 0.1320 (43.3\%) & 0.1797 (37.8\%) & 0.1160 (45.5\%) & 0.1305 (42.0\%) \\
    \bottomrule
    &       &       & \multicolumn{1}{r}{} &       &       &       & \multicolumn{1}{r}{} &       &       &       &  \\
    \toprule
    \multicolumn{4}{c|}{Component} & \multicolumn{4}{c|}{Gowalla}  & \multicolumn{4}{c}{Yelp2022} \\
    \midrule
    HS    & PI    & HA    & GP    & Recall@10  & Recall@20  & NDCG@10  & NDCG@20  & Recall@10  & Recall@20  & NDCG@10  & NDCG@20 \\
    \midrule
          &       &       &       & 0.1224 (0.0\%) & 0.1815 (0.0\%) & 0.1172 (0.0\%) & 0.1348 (0.0\%) & 0.0641 (0.0\%) & 0.1082 (0.0\%) & 0.0536 (0.0\%) & 0.0678 (0.0\%) \\
    \checkmark     &       &       &       & 0.1266 (3.4\%) & 0.1843 (1.5\%) & 0.1228 (4.8\%) & 0.1393 (3.3\%) & 0.0697 (8.7\%) & 0.1148 (6.1\%) & 0.0584 (9.0\%) & 0.0728 (7.4\%) \\
          & \checkmark     & \checkmark     &       & 0.1330 (8.7\%) & 0.1974 (8.8\%) & 0.1284 (9.6\%) & 0.1475 (9.4\%) & 0.0715 (11.5\%) & 0.1179 (9.0\%) & 0.0597 (11.4\%) & 0.0746 (10.0\%) \\
    \checkmark     &       & \checkmark     &       & 0.1376 (12.4\%) & 0.1981 (9.1\%) & 0.1338 (14.2\%) & 0.1512 (12.2\%) & 0.0769 (20.0\%) & 0.1242 (14.8\%) & 0.0644 (20.1\%) & 0.0795 (17.3\%) \\
    \checkmark     & \checkmark     &       &       & 0.1411 (15.3\%) & 0.2083 (14.8\%) & 0.1354 (15.5\%) & 0.1550 (15.0\%) & 0.0775 (20.9\%) & 0.1247 (15.2\%) & 0.0650 (21.3\%) & 0.0800 (18.0\%) \\
    \checkmark     & \checkmark     & \checkmark     &       & 0.1447 (18.2\%) & 0.2086 (14.9\%) & 0.1396 (19.1\%) & 0.1581 (17.3\%) & 0.0790 (23.2\%) & 0.1273 (17.7\%) & 0.0660 (23.1\%) & 0.0814 (20.1\%) \\
    \checkmark     & \checkmark     & \checkmark     & \checkmark     & 0.1576 (28.8\%) & 0.2273 (25.2\%) & 0.1532 (30.7\%) & 0.1734 (28.6\%) & 0.0795 (24.0\%) & 0.1280 (18.3\%) & 0.0670 (25.0\%) & 0.0824 (21.5\%) \\
    \bottomrule
    \end{tabular}}%
  \label{tab:abla1}%
\end{table*}%

\subsection{Performance Comparison (RQ1)}

Table \ref{tab:main} shows the test set results, highlighting the best (bold), second-best (underline), and relative gains ($\Delta$ and $\Delta_+$) of HGCH and HGCH+ over the strongest baseline. From the table, we observe:

\begin{itemize}[leftmargin=*]
\item HGCH outperforms all baselines when only interaction data is considered, with up to 18.06\% improvement. Hyperbolic models, including HGCH, HGCF, HRCF, and HICF, have an edge over Euclidean models for modeling large user-item networks with power-law distributions. Among hyperbolic models, HGCH benefits more from data quantity. For example, HGCH improves more on larger datasets like Amazon-Book~(+14.74\% for Recall@10 and +18.06\% for NDCG@10) than on Amazon-CD~(+3.49\% for Recall@10 and +4.27\% for NDCG@10) compared to second-best HICF, indicating better use of hyperbolic space.

\item The incorporation of side information into our model, resulting in HGCH+, yields further enhancements across all datasets. In contrast, other hyperbolic models' performance varies depending on the dataset; they sometimes exhibit significant improvements, while at other times their performance remains stagnant or even declines. A case in point is the Yelp2022 dataset, where other hyperbolic models incorporating side information suffer a considerable performance dip, while our HGCH+ maintains a modest improvement. This highlights the potential pitfalls of direct HCG usage, while validating HGCH's more effective utilization of side information.
\end{itemize}

\subsection{Ablation Analysis (RQ2)}
\label{sec:aa}

To further test our proposal, we evaluated how each HGCH component affects the performance. Table \ref{tab:abla1} show the results. Here, HS, PI, HA, and GP mean Hyperbolic negative Sampling, Power law prior-based Initialization, Hyperbolic neighbor Aggregation, and Gate fusion with Prior, respectively. Parentheses show the improvement over the first row (Base) of the table. Table \ref{tab:abla1} reveals that:
\begin{itemize}[leftmargin=*]
\item Our HS enhances performance on each dataset (row 2 in Table \ref{tab:abla1}), and this improvement is strongly correlated with the number of items in each dataset. The ranking of the improvements is consistent with the ranking of the number of items in each dataset.
\item Coupling PI with HS results in further performance boosts across all datasets (row 5 in Table \ref{tab:abla1}), especially for Amazon-Book, where improvement peaks at 39.4\%. This highlights the importance of initialization for hyperbolic models, as our PI can provide more informative prior knowledge for the initial embedding.
\item Incorporating HA with HS likewise yields performance enhancements across all datasets (row 4 in Table \ref{tab:abla1}). However, when combined with PI, the performance (row 6 in Table \ref{tab:abla1}) varies among datasets compared to row 5 in Table \ref{tab:abla1}: remaining unchanged for Amazon-CD and Amazon-Book, while seeing a boost for Gowalla and Yelp2022 datasets. This seems to correlate with dataset types, namely purchase and check-in datasets respectively, suggesting that the efficacy of HA may be dataset-type dependent. In our experiments, HA's impact is more evident in the check-in datasets.
\item HS and PI+HA are complementary, as their combination produces an improvement equivalent to 1+2$>$3 in metrics (refer to rows 2, 3, and 6 in Table \ref{tab:abla1}). This can be attributed to their distinct model enhancement aspects, namely model training, model initialization, and model structure respectively.
\item Upon introducing GP (HGCH+) (row 7 in Table \ref{tab:abla1}), the model performance is further enhanced, particularly for Amazon-CD, which sees the highest further improvement rate of 13.9\% compared to HGCH (row 6 in Table \ref{tab:abla1}). This verifies the effectiveness of GP and side information. However, the improvement for Yelp2022 is less pronounced, possibly due to the limited utility and noises of its side information in the recommendation task.
\end{itemize}

\begin{figure*}[!t]
\centering
\includegraphics[width=1.7in]{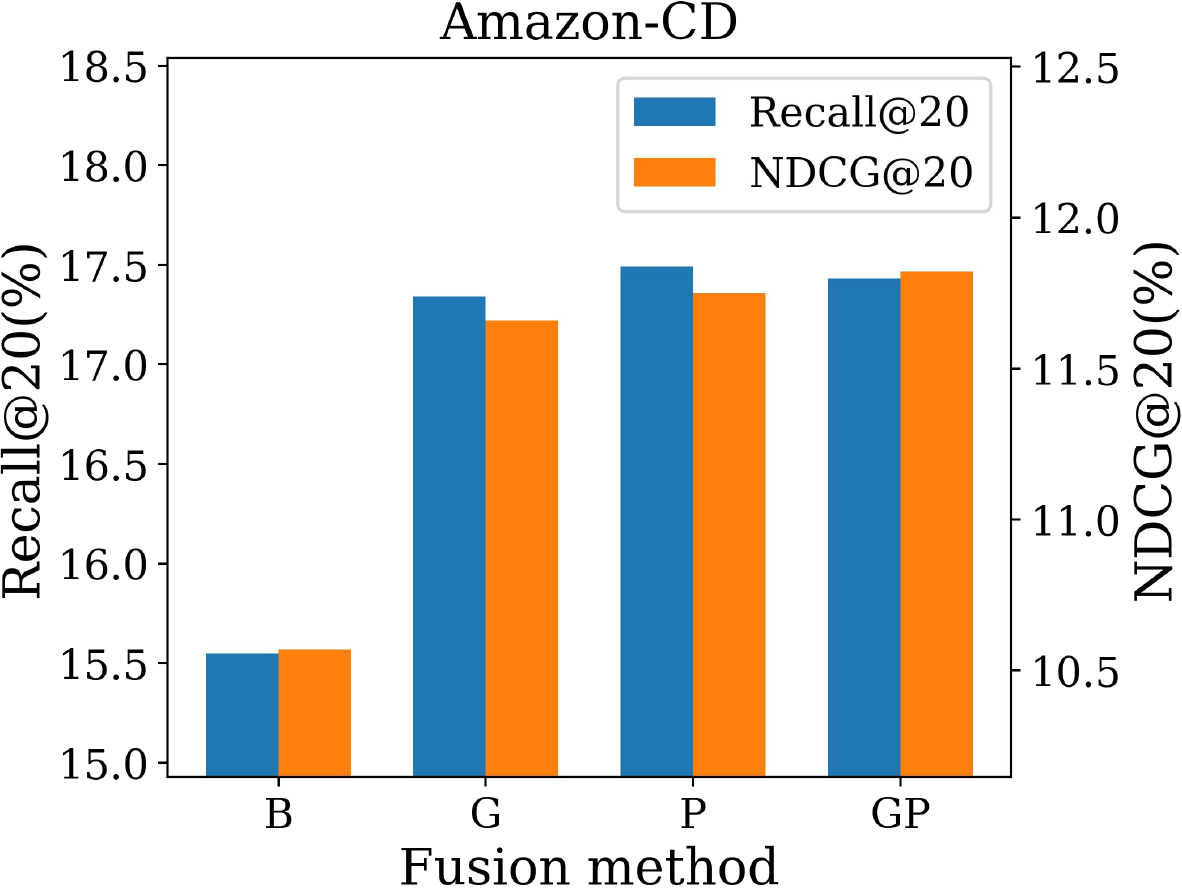}
\includegraphics[width=1.7in]{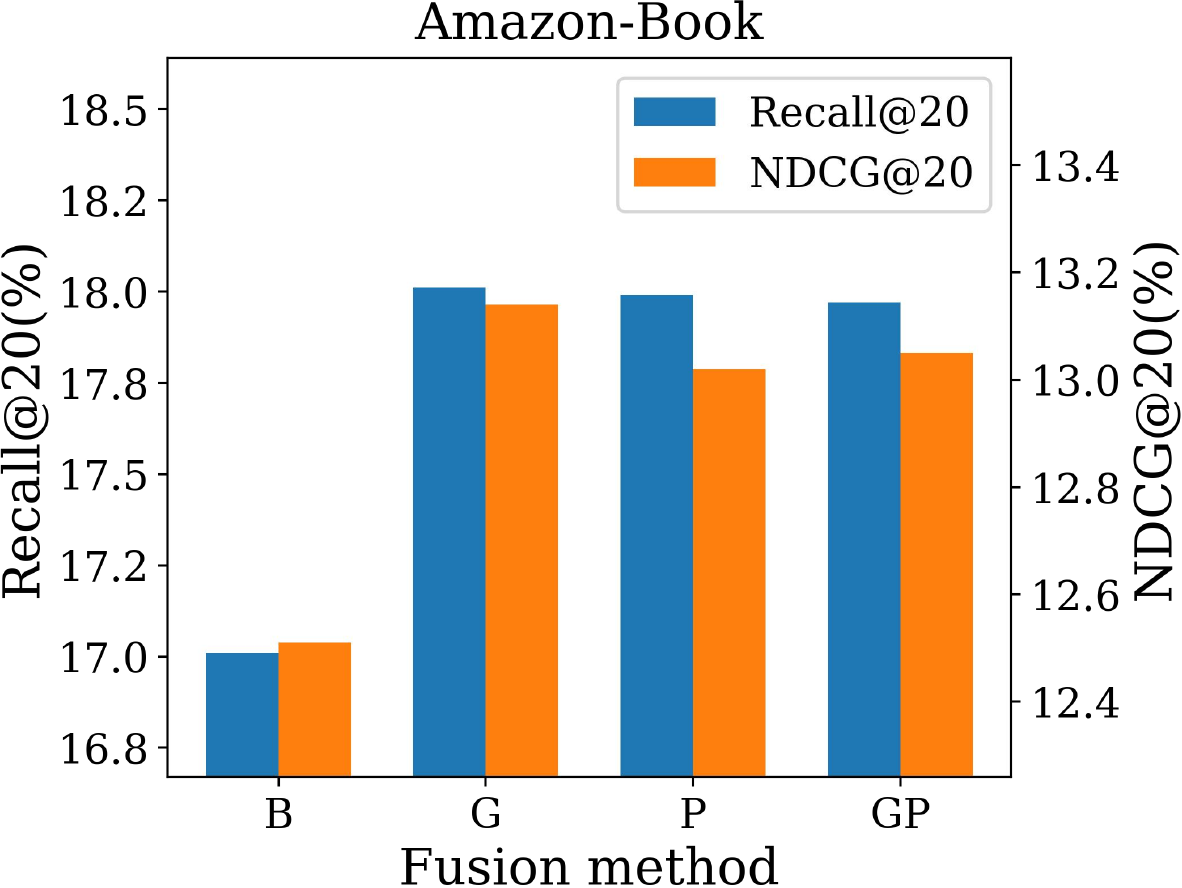}
\includegraphics[width=1.7in]{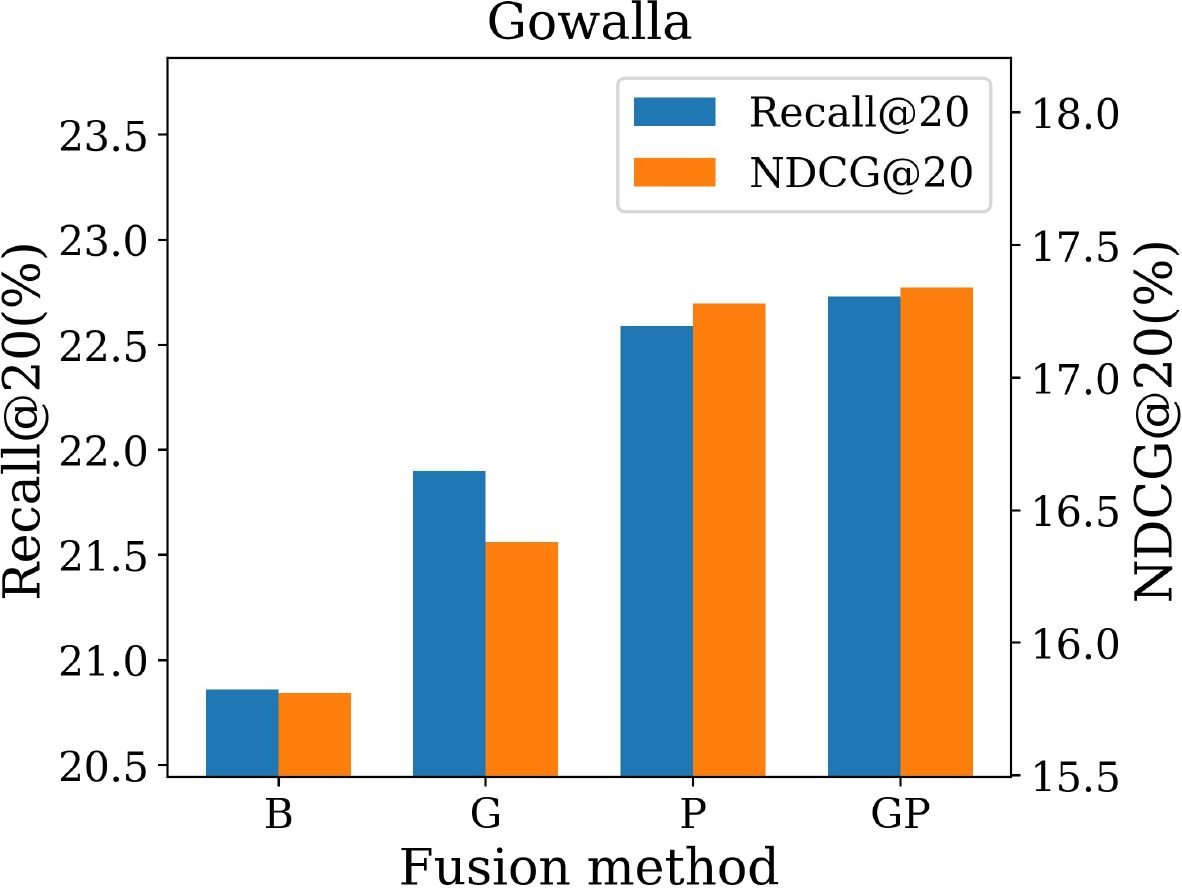}
\includegraphics[width=1.7in]{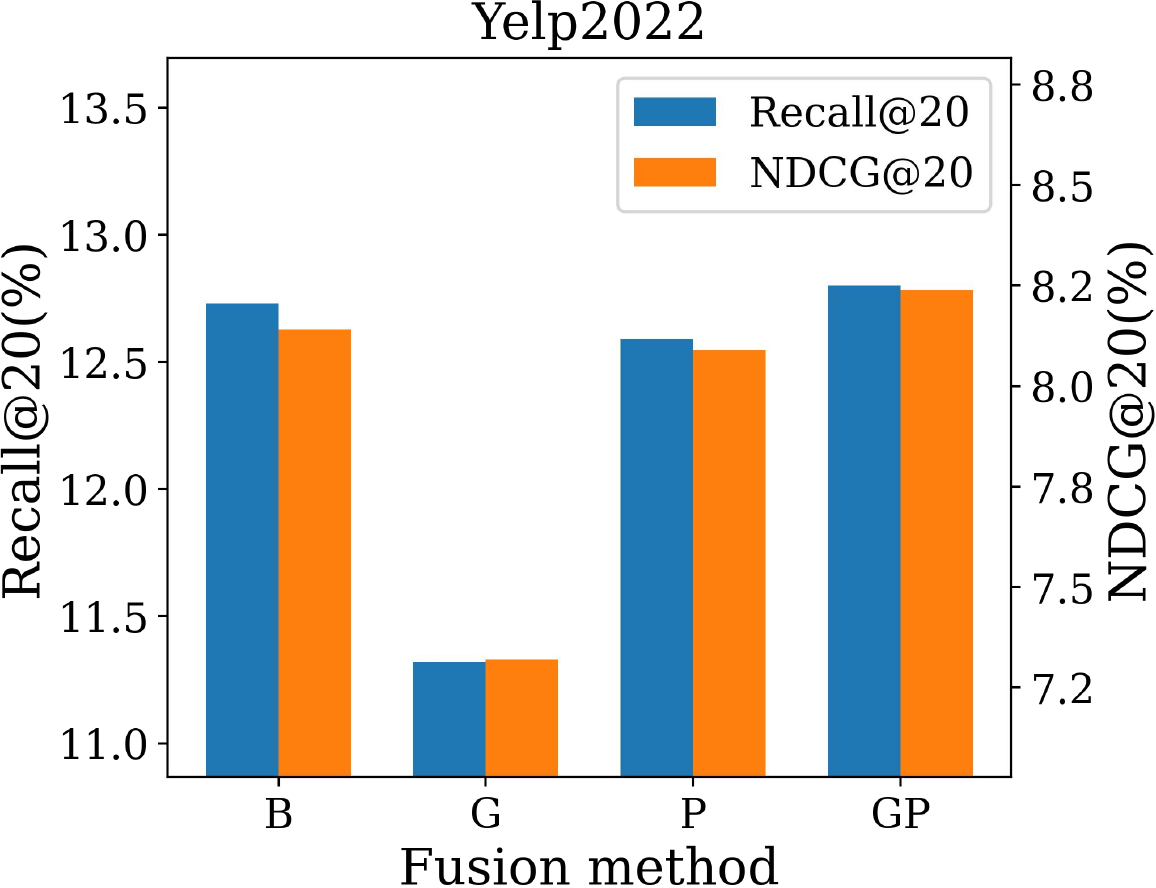}
\caption{Comparison between different fusion methods. B, G, P, and GP represent Base, Gate, Prior, and Gate\&Prior, respectively, where the Base is the HGCH.}
\label{fig:fusion}
\end{figure*}

\begin{figure*}[!t]
\centering
\subfloat[Item embeddings \textbf{before} the graph convolution layers]{
\includegraphics[width=1.87in]{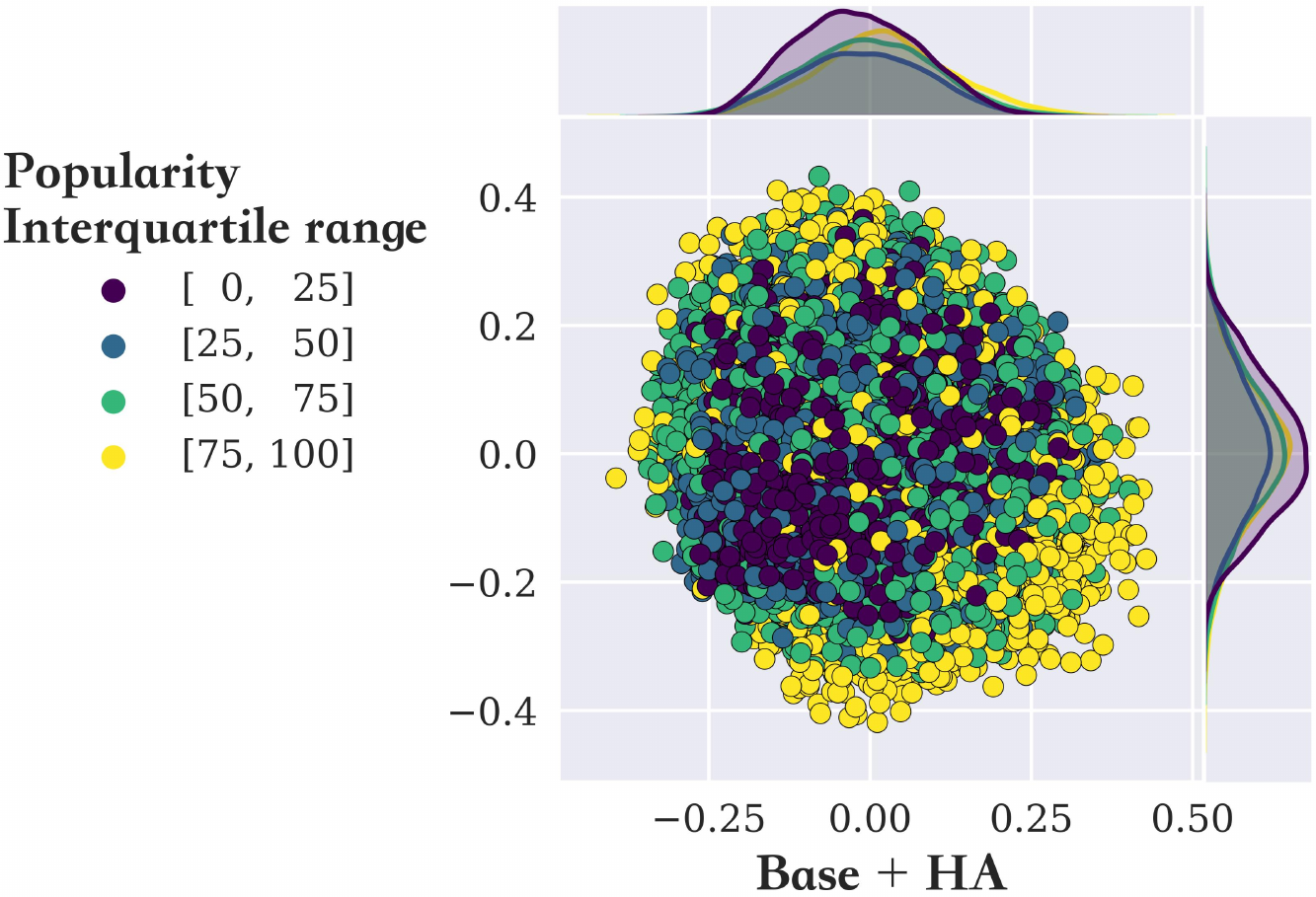}
\hspace{1.25cm}
\includegraphics[width=1.26in]{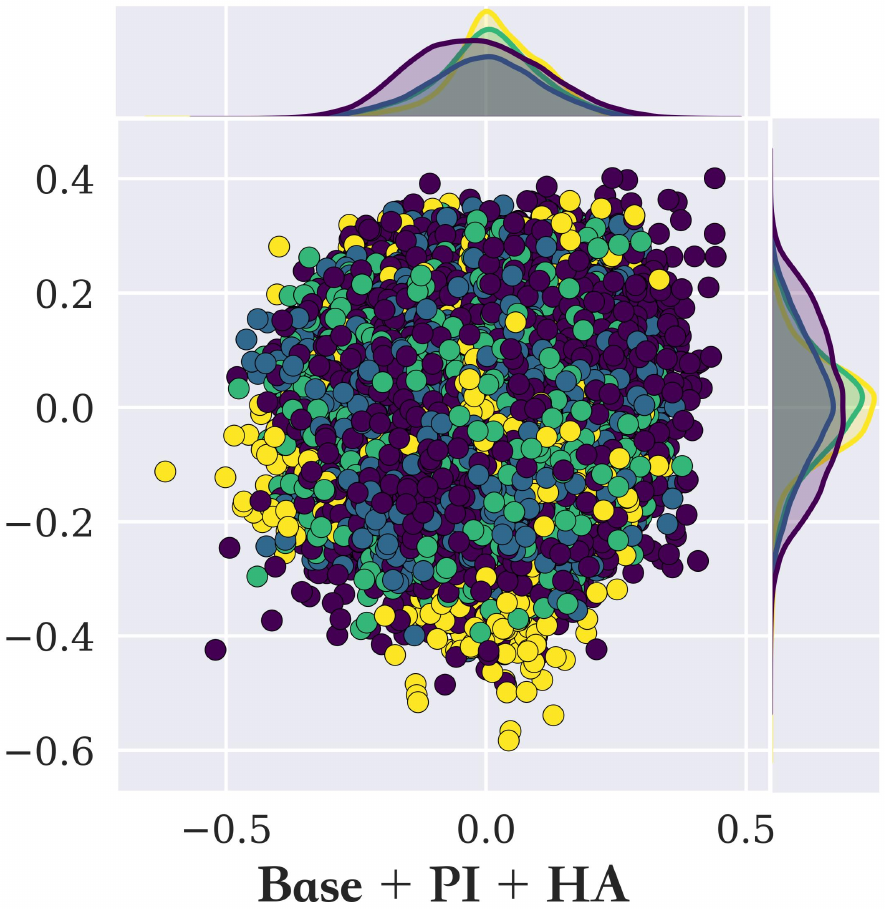}
\hspace{1.25cm}
\includegraphics[width=1.26in]{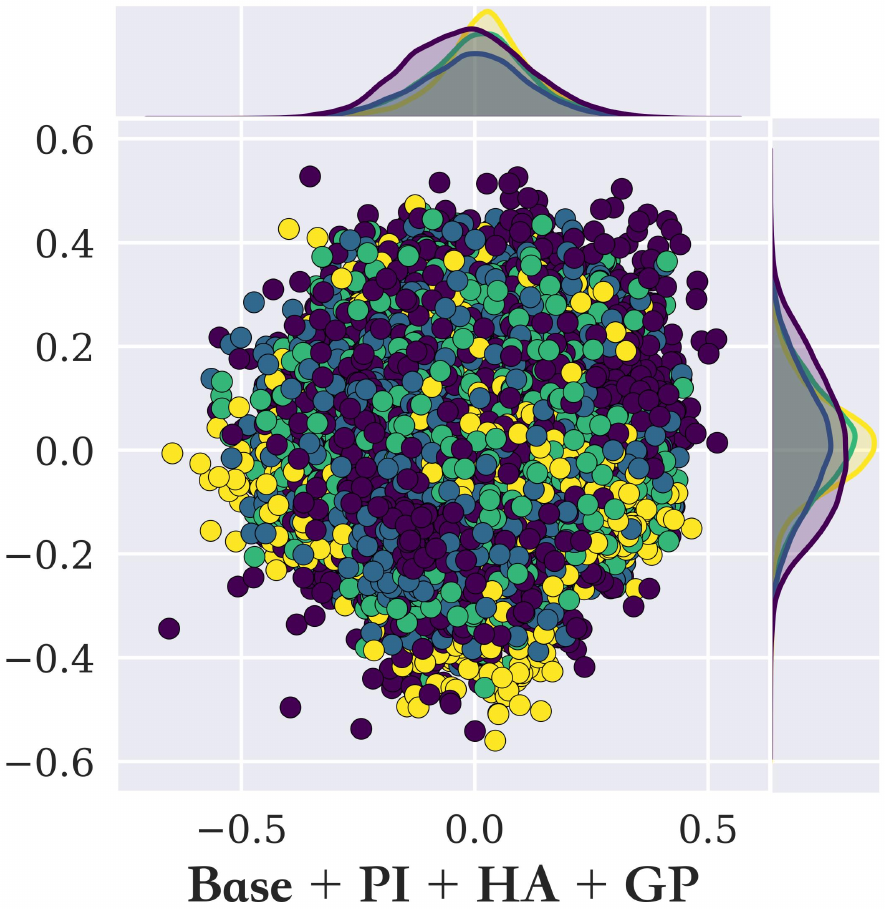}
\label{fig:before}}
\hfil
\subfloat[Item embeddings \textbf{after} the graph convolution layers]{
\includegraphics[width=1.87in]{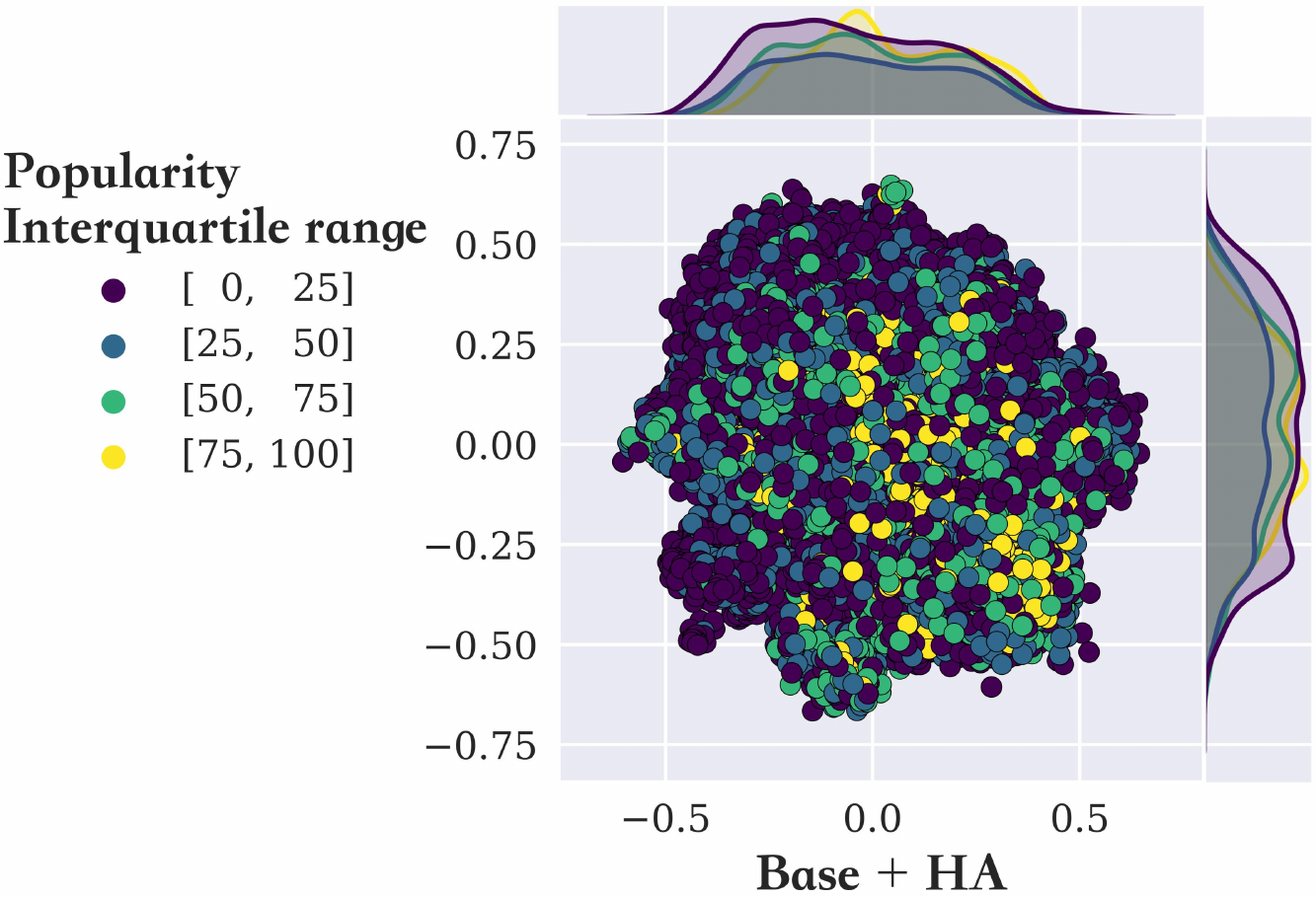}
\hspace{1.25cm}
\includegraphics[width=1.27in]{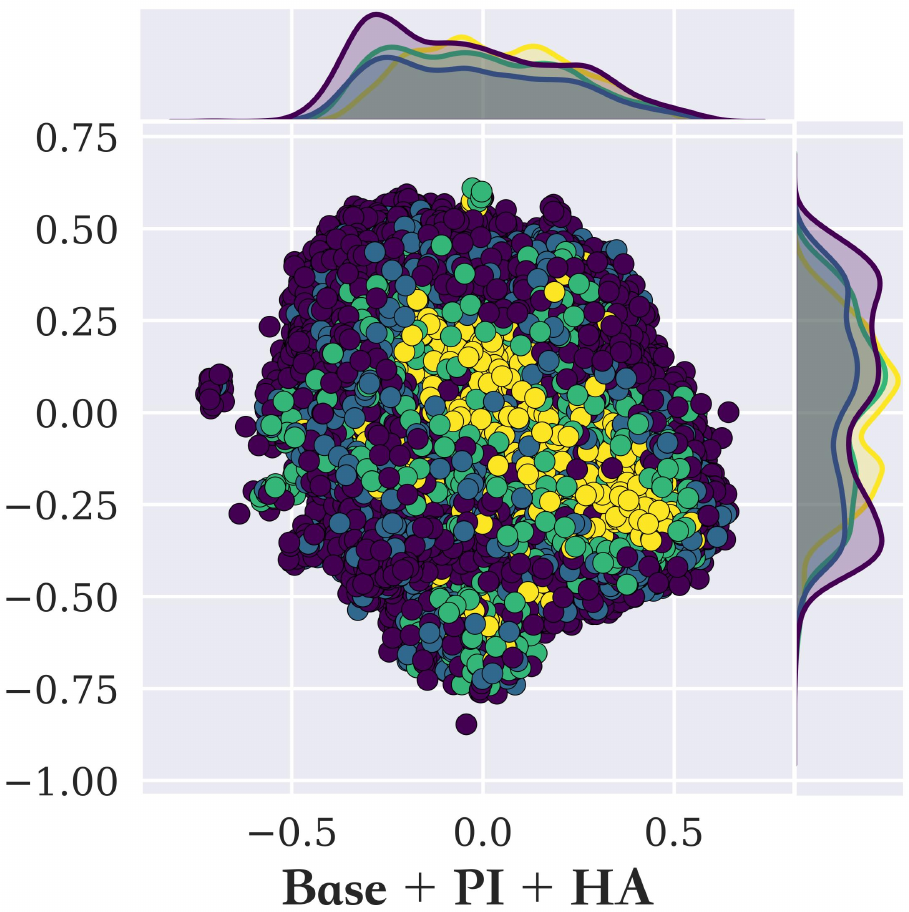}
\hspace{1.25cm}
\includegraphics[width=1.27in]{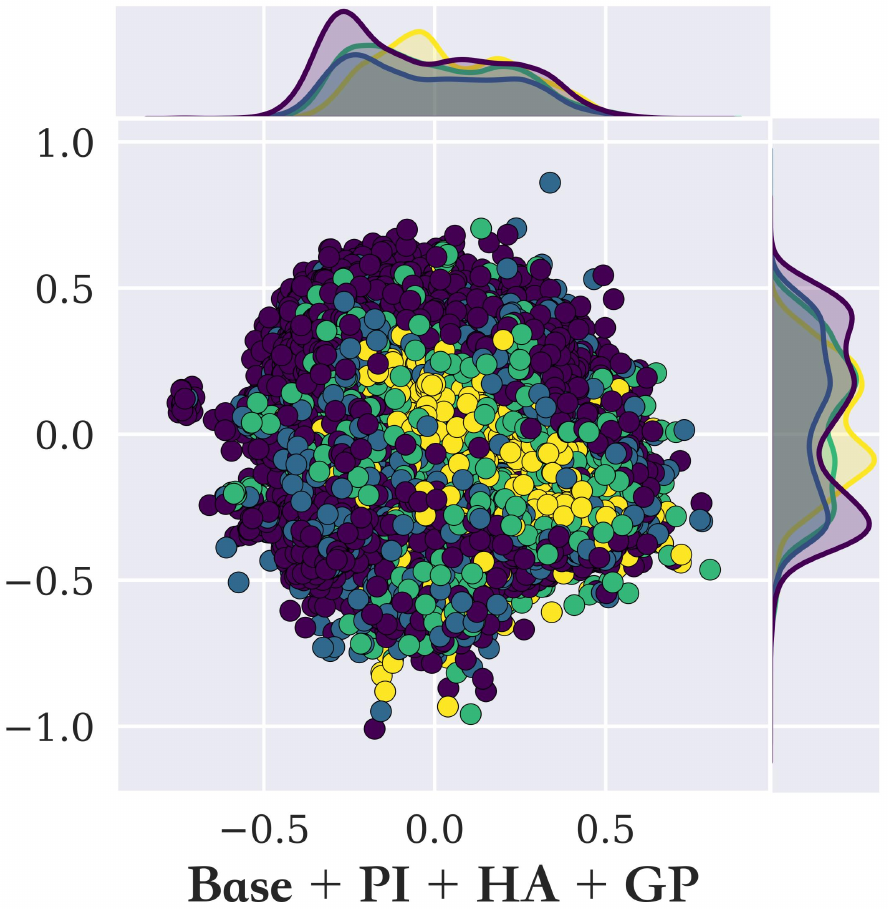}
\label{fig:after}}
\caption{HGCH item embedding visualization in the Poincaré representation of tangent space before and after the graph convolutional layers on the dataset Amazon-Book. Items are categorized into quartiles by popularity, where [75, 100] represents the most popular items. The meaning of Base, PI, HA, and GP are the same as in Section \ref{sec:aa}.}
\label{fig:visual}
\end{figure*}

\subsection{Comparison of Fusion Methods (RQ3)}
\label{sec:cofm}
We conducted an empirical evaluation to assess the impact of different fusion methods on the model performance. Fig. \ref{fig:fusion} illustrates the results of the fusion methods that we introduced in Section \ref{sec:msif}, where the Base is the version without side information, and B, G, P, and GP denote Base, Gate, Prior, and Gate\&Prior, respectively. We observe two salient phenomena:
\begin{itemize}[leftmargin=*]
\item Except for gate in the Yelp2022 dataset, all three fusion methods bring performance improvement compared to the Base, indicating that these fusion methods basically incorporate side information into embedding.
\item The Gate and Prior methods' efficacy depends on specific factors. Gate excels in large datasets, outperforming other methods on the Amazon-Book dataset. Conversely, Prior maintains good performance in most cases. However, both struggle with high noise levels in side information, demonstrated by their decreased performance in Yelp2022's noisy data.\footnote{Yelp2022's noise in side information also significantly impacts HGCF, HRCF, and HICF, as shown in Table \ref{tab:main}.} Under such conditions, their efficiency falls below Base. The combined Gate\&Prior method shows stable improvement across datasets, highlighting its capability to adjust weights effectively based on Prior.
\end{itemize}

\subsection{Embedding Visualization Analysis (RQ4)}

In this section, we aim to provide a plausible interpretation of our proposed approach by visualizing the embedding learned by HGCH. Since our model employs the Poincaré disc as its underlying geometry, it facilitates the visualization process. We adopt a 3-layer HGCH model with an embedding size of 2 for visualization purposes, following the previous work \cite{HME,HGCF}. To comprehend the learned embeddings, we select the Amazon-Book dataset as an exemplar, assign the quartiles of the number of interactions as node labels with different colors, and display the embeddings before and after graph convolution.

As showed in Fig. \ref{fig:visual}, before graph convolution, with uniform initialization, item embeddings show a Gaussian-like spread. After applying power law prior-based initialization and then adding side information, popular items cluster near the origin, indicating enhanced embedding training. After graph convolution, embeddings with power law prior-based initialization exhibit clearer class boundaries compared to uniform initialization. Moreover, integrating side information further congregates popular nodes near the origin, suggesting its effectiveness in performance improvement by aligning node distribution with popularity.

\subsection{Convergent Speed (RQ5)}
We also examine the convergence behavior of our proposed HGCH model compared to other hyperbolic models throughout the training process. Fig. \ref{fig:con} shows the NDCG@10 metrics on the validation set from 1 to 1000 epochs. Similar patterns are observed for other metrics@$K$s. Notably, HGCH demonstrates exceptionally rapid convergence, allowing us to limit its training to 40 epochs on Amazon-Book and 60 epochs on Yelp2022, thereby saving computational resources. Additionally, on Yelp, HGCF shows no improvement in NDCG@10 during the last 100 epochs of training, leading to early termination. From these observations, we can draw the following conclusions: (1) our HGCH consistently outperforms all baseline models across all epochs, showing significant improvement; and (2) our approach achieves peak performance in fewer epochs, indicating that our hyperbolic sampling method accelerates the training process effectively.

\section{Conclusion}
\label{sec:conclu}

In this paper, we propose HGCH, a HGCN-based collaborative filtering model that improves the HGCN structure and incorporates multiple kinds of side information into a heterogeneous collaborative graph, as well as greatly speeds up the training convergence speed.
Experiments conducted on four real-world datasets showed significant improvements compared to the baseline and better utilization of hyperbolic space (as evidenced by better utilization of exponentially increased capacity to pay more attention to tail items), which is beneficial for personalized recommendations and increasing market diversity \cite{HICF}. We believe it will bring new insights into HGCN-based recommender systems.

\bibliographystyle{ACM-Reference-Format}
\balance
\bibliography{ref.bib}

\clearpage
\appendix

\section{Anaysis of Gyromidpoint}
\label{sec:aog}
Here, we try to analyze more specifically how the neighbor aggregation through gyromidpoint differs from the way by tangent space, to explain why hyperbolic neighbor aggregation may be more effective. Substituting Eq. (10) into $\lambda({k},{\mathbf{h}_i})$ and omit the notation of subspace and the number of layer for simplicity, we get:
\begin{align*}
\lambda(k,{\mathbf{h}_i})&=2\left(1-\frac{\|\mathbf{h}_i\|^{2}}{k}\right)^{-1}\nonumber \\
&=2\left(1-\frac{\|\exp _\mathbf{o}^{k}(\mathbf{e}_i)\|^{2}}{k}\right)^{-1}\nonumber \\
&=2\left(1-\frac{\|\tanh \left(\frac{\|\mathbf{e}_i\|}{\sqrt{k}}\right) \frac{\sqrt{k}\mathbf{e}_i}{\|\mathbf{e}_i\|}\|^{2}}{k}\right)^{-1}\nonumber \\
&=2\left(1-\text{tanh}^2\left(\frac{\|\mathbf{e}_{i}\|}{\sqrt{k}}\right)\right)^{-1}.
\end{align*}
We notice that when fixing the negative reciprocal of the curvature $k$, the magnitude of $\lambda(k,{\mathbf{h}_i})$ positively correlates with the magnitude of $\mathbf{e}_i$. Furthermore, the role of $\lambda(k,{\mathbf{h}_i})$ in Eq. (9) is similar to the weight of node $\mathbf{h}_i$, which means that the node with a larger embedding magnitude in tangent space has a more significant weight in neighbor aggregation. Therefore, combining the relationship between embedding magnitude and popularity in hyperbolic space, we can find that: in gyromidpoint neighbor aggregation, nodes with lower popularity are more likely to contribute more. While tangent space neighbor aggregation only directly averages all the nodes being aggregated, and its corresponding Laplace matrix is $\mathbf{D}^{-1}\mathbf{A}$ ($\mathbf{A}$ is the adjacency matrix, and $\mathbf{D}$ is the degree matrix), which does not explicitly weaken the contribution of high popularity nodes, thus may lead to the degradation of performance. There is also a commonly used Laplacian matrix of $\mathbf{D}^{-\frac{1}{2}}\mathbf{A}\mathbf{D}^{-\frac{1}{2}}$, which does a second averaging according to the popularity of the nodes being aggregated, but in practice, it yields a worse performance under the hyperbolic space scenario.

\section{DATASETS AND BASELINES}

\subsection{Dataset Description}
In this section, we give the detailed descriptions and pre-processing
of the four datasets.

\noindent
\textbf{Amazon-CD} and \textbf{Amazon-Book}: Amazon-review is a widely used dataset for product recommendations \cite{rec}. Following previous work \cite{HGCF}, we select Amazon-CD and Amazon-Book from this set and transform the ratings into binary preferences with a threshold >=4 to simulate implicit feedback. To ensure the data quality, we use a 10-core setting and a 5-core setting for Amazon-CD for users and items, respectively, \textit{i.e.}, retaining users with at least ten interactions and items with five interactions. Moreover, a 10-core setting is used for Amazon-Book.

\noindent
\textbf{Gowalla}: This is a dataset collected from gowalla, a location-based social networking website where users share their locations by checking in \cite{Gowalla}. Here, we consider the location as the item. We also use a 10-core to ensure data quality.

\noindent
\textbf{Yelp2022}: This dataset is adopted from the 2022 edition of the yelp website. Here, we consider local businesses like restaurants and bars as the item. Again, we use a 10-core setting to ensure that the user and item have at least ten interactions.

\begin{table}[!t]
  \centering
  \caption{Interaction information statistics for datasets.}
  \setlength{\tabcolsep}{1.8mm}{
    \begin{tabular}{l|cccc}
    \toprule
    \textbf{Dataset} & \textbf{\#User} & \textbf{\#Item} & \textbf{\#Interaction} & \textbf{Density} \\
    \midrule
    \textbf{Amazon-CD} & 16865  & 33900  & 476676  & 0.083\% \\
    \textbf{Amazon-Book} & 109730  & 96421  & 3181759  & 0.030\% \\
    \textbf{Gowalla} & 29858  & 40988  & 1027464  & 0.084\% \\
    \textbf{Yelp2022} & 93537  & 53347  & 2533759  & 0.051\% \\
    \bottomrule
    \end{tabular}%
    }
  \label{tab:inter_stat}%
\end{table}%

\begin{table}[!t]
  \centering
  \caption{Side information statistics for datasets. U, I, and C represent user, item, and category respectively.}
  \setlength{\tabcolsep}{1.8mm}{
    \begin{tabular}{l|cccc}
    \toprule
    \textbf{Dataset} & \textbf{Relation (A-B)} & \textbf{\#A} & \textbf{\#B} & \textbf{\#A-B} \\
    \midrule
    \multirow{3}[1]{*}{\textbf{Amazon-CD}} & category (C-I) & 380   & 33900  & 68382 \\
          & also\_buy (I-I) & 33900  & 33900  & 121149 \\
          & also\_view (I-I) & 33900  & 33900  & 121149 \\
    \multirow{3}[0]{*}{\textbf{Amazon-Book}} & category (C-I) & 457   & 96421  & 171257 \\
          & also\_buy (I-I) & 96421 & 96421 & 347311 \\
          & also\_view (I-I) & 96421 & 96421 & 159154 \\
    \multirow{2}[0]{*}{\textbf{Gowalla}} & friend (U-U) & 29858 & 29858 & 279478 \\
          & neighbor (I-I) & 40988 & 40988 & 501242 \\
    \multirow{2}[1]{*}{\textbf{Yelp2022}} & friend (U-U) & 93537  & 93537  & 1295854 \\
          & neighbor (I-I) & 53347  & 53347  & 1038744 \\
    \bottomrule
    \end{tabular}%
    }
  \label{tab:side_stat}%
\end{table}%

Moreover, we need to build side information bipartite graphs (referred to as side graphs for short) for each dataset. For Amazon-CD and Amazon-Book, we choose $\{$category (C-C), also$\_$buy (I-I), also$\_$view (I-I)$\}$. For Gowalla and Yelp2022, we first choose social information (U-U) and then construct geographic neighborhood information (I-I) from latitude and longitude, which is an important feature in poi recommendation. We approximate the earth as a sphere with a radius of 6371 km and use the haversine\footnote{\href{https://en.wikipedia.org/wiki/Haversine\_formula}{https://en.wikipedia.org/wiki/Haversine\_formula.}} function to calculate the geographical distance $d_{geo}$ between two locations. Locations within a geographical distance of 0.2 km are considered neighbors and form the geographical neighbor graph.

We present the statistics of historical interaction data and side information for the four datasets in Table \ref{tab:inter_stat} and Table \ref{tab:side_stat}. For each dataset, we divide it into a training set and a test set with the ratio of 8:2. Then, for tuning hyperparameters, 10$\%$ of the training set is randomly selected as the validation set.

\subsection{Baselines and Settings}
This section gives clear descriptions and settings of the baselines
used in our experiments.

\begin{table*}[!htb]
\footnotesize
  \centering
  \caption{Performance of HGCH with different subsets of components on the H20 and T80 items for dataset Amazon-Book and Yelp2022.}
  \setlength{\tabcolsep}{1.3mm}{
    \begin{tabular}{cccc|cc|cc|cc|cc}
    \toprule
    \multicolumn{4}{c|}{\multirow{2}[4]{*}[+1.5ex]{Component}} & \multicolumn{8}{c}{Amazon-Book} \\
\cmidrule{5-12}    \multicolumn{4}{c|}{}         & \multicolumn{2}{c|}{Recall@20} & \multicolumn{2}{c|}{Recall@10} & \multicolumn{2}{c|}{NDCG@20} & \multicolumn{2}{c}{NDCG@10} \\
\cmidrule{1-4}    HS    & PI    & HA    & GP    & H20   & T80   & H20   & T80   & H20   & T80   & H20   & T80 \\
    \midrule
          &       &       &       & 0.1143(0.0\%) & 0.0445(0.0\%) & 0.0813(0.0\%) & 0.0315(0.0\%) & 0.0819(0.0\%) & 0.0325(0.0\%) & 0.0717(0.0\%) & 0.0285(0.0\%) \\
    \checkmark     &       &       &       & 0.1232(7.8\%) & 0.0490(10.1\%) & 0.0922(13.4\%) & 0.0368(16.8\%) & 0.0937(14.4\%) & 0.0376(15.7\%) & 0.0847(18.1\%) & 0.0340(19.3\%) \\
          & \checkmark     & \checkmark     &       & 0.1247(9.1\%) & 0.0512(15.1\%) & 0.0889(9.3\%) & 0.0368(16.8\%) & 0.0894(9.2\%) & 0.0378(16.3\%) & 0.0784(9.3\%) & 0.0333(16.8\%) \\
    \checkmark     &       & \checkmark     &       & 0.1399(22.4\%) & 0.0600(34.8\%) & 0.1049(29.0\%) & 0.0450(42.9\%) & 0.1059(29.3\%) & 0.0458(40.9\%) & 0.0957(33.5\%) & 0.0414(45.3\%) \\
    \checkmark     & \checkmark     &       &       & 0.1423(24.5\%) & 0.0616(38.4\%) & 0.1063(30.8\%) & 0.0463(47.0\%) & 0.1076(31.4\%) & 0.0469(44.3\%) & 0.0972(35.6\%) & 0.0424(48.8\%) \\
    \checkmark     & \checkmark     & \checkmark     &       & 0.1418(24.1\%) & 0.0610(37.1\%) & 0.1070(31.6\%) & 0.0459(45.7\%) & 0.1078(31.6\%) & 0.0470(44.6\%) & 0.0978(36.4\%) & 0.0426(49.5\%) \\
    \checkmark     & \checkmark     & \checkmark     & \checkmark     & 0.1458(27.6\%) & 0.0698(56.9\%) & 0.1093(34.4\%) & 0.0524(66.3\%) & 0.1102(34.6\%) & 0.0529(62.8\%) & 0.0997(39.1\%) & 0.0479(68.1\%) \\
    \midrule
          &       &       & \multicolumn{1}{r}{} &       & \multicolumn{1}{r}{} &       & \multicolumn{1}{r}{} &       & \multicolumn{1}{r}{} &       &  \\
    \midrule
    \multicolumn{4}{c|}{\multirow{2}[4]{*}[+1.5ex]{Component}} & \multicolumn{8}{c}{Yelp2022} \\
\cmidrule{5-12}    \multicolumn{4}{c|}{}         & \multicolumn{2}{c|}{Recall@20} & \multicolumn{2}{c|}{Recall@10} & \multicolumn{2}{c|}{NDCG@20} & \multicolumn{2}{c}{NDCG@10} \\
\cmidrule{1-4}    HS    & PI    & HA    & GP    & H20   & T80   & H20   & T80   & H20   & T80   & H20   & T80 \\
    \midrule
          &       &       &       & 0.1087(0.0\%) & 0.0365(0.0\%) & 0.0660(0.0\%) & 0.0217(0.0\%) & 0.0688(0.0\%) & 0.0234(0.0\%) & 0.0551(0.0\%) & 0.0186(0.0\%) \\
    \checkmark     &       &       &       & 0.1156(6.3\%) & 0.0377(3.3\%) & 0.0718(8.8\%) & 0.0231(6.5\%) & 0.0741(7.7\%) & 0.0247(5.6\%) & 0.0603(9.4\%) & 0.0201(8.1\%) \\
          & \checkmark     & \checkmark     &       & 0.1166(7.3\%) & 0.0384(5.2\%) & 0.0719(8.9\%) & 0.0232(6.9\%) & 0.0742(7.8\%) & 0.0249(6.4\%) & 0.0600(8.9\%) & 0.0201(8.1\%) \\
    \checkmark     &       & \checkmark     &       & 0.1236(13.7\%) & 0.0411(12.6\%) & 0.0782(18.5\%) & 0.0253(16.6\%) & 0.0800(16.3\%) & 0.0270(15.4\%) & 0.0656(19.1\%) & 0.0220(18.3\%) \\
    \checkmark     & \checkmark     &       &       & 0.1239(14.0\%) & 0.0429(17.5\%) & 0.0783(18.6\%) & 0.0266(22.6\%) & 0.0803(16.7\%) & 0.0281(20.1\%) & 0.0659(19.6\%) & 0.0229(23.1\%) \\
    \checkmark     & \checkmark     & \checkmark     &       & 0.1257(15.6\%) & 0.0439(20.3\%) & 0.0797(20.8\%) & 0.0271(24.9\%) & 0.0814(18.3\%) & 0.0286(22.2\%) & 0.0668(21.2\%) & 0.0233(25.3\%) \\
    \checkmark     & \checkmark     & \checkmark     & \checkmark     & 0.1266(16.5\%) & 0.0453(24.1\%) & 0.0803(21.7\%) & 0.0280(29.0\%) & 0.0824(19.8\%) & 0.0297(26.9\%) & 0.0678(23.0\%) & 0.0243(30.6\%) \\
    \bottomrule
    \end{tabular}}%
  \label{tab:abla2}%
\end{table*}%

To show the effectiveness of our model, we compare our proposed HGCH with latent factor-based (BPRMF \cite{BPR} and WRMF \cite{WRMF}), graph neural network-based (NGCF \cite{NGCF}, LightGCN \cite{LightGCN}, and DGCF \cite{DGCF}), and hyperbolic neural network-based (HGCF \cite{HGCF}, HRCF \cite{HRCF}, and HICF \cite{HICF}):
\begin{itemize}[leftmargin=*]
\item \textbf{BPRMF} \cite{BPR} is based on Bayesian personalized ranking, which models the order of candidate items by pairwise ranking loss.
\item \textbf{WRMF} \cite{WRMF} is a classical latent factor model that attaches a weight to each training sample to characterize the confidence level of the user's preference for the items.
\item \textbf{NGCF} \cite{NGCF} is a GCN-based model that explores high-order connectivity in the user-item graph by propagating embedding and injects collaboration signals into the embedding explicitly.
\item \textbf{LightGCN} \cite{LightGCN} is a state-of-the-art GCN model, which removes feature transformations and nonlinear activations that are useless for collaborative filtering in GCN.
\item \textbf{DGCF} \cite{DGCF} uses GCN to encode graph structure data and iteratively refines intention-aware interaction graphs and representations by modeling the intention distribution of each user-item interaction. Here, we follow the original paper to set its graph disentangling layer to 1 because of high computational cost.
\item \textbf{HGCF} \cite{HGCF} is a leading hyperbolic GCN model, which replaces the traditional Euclidean space with Lorentz space, and learns embedding by Skip-GCN structure and margin ranking loss. 
\item \textbf{HRCF} \cite{HRCF} is a leading hyperbolic GCN model, which is based on HGCF and adds a geometry-aware hyperbolic regularizer to boost optimization. 
\item \textbf{HICF} \cite{HICF} is a state-of-the-art hyperbolic GCN model, which is also based on HGCF and adapt the hyperbolic margin ranking learning,
making its pull and push procedure geometric-aware, then providing informative guidance for the learning of both head and tail items.
\end{itemize}
Besides, we refer to the version using HCG directly as HGCF+, HRCF+, and HICF+, respectively. For our model, we denote the version that uses only the user-item bipartite graph and the version that utilizes HCG as HGCH and HGCH+, respectively. Note that the main difference between HGCF+/HRCF+/HICF+ and HGCH+ is the neighbor aggregation, the former proceeds directly on the HCG, but the latter performs on each subgraph of the HCG separately.

We used RecBole \cite{recbole1.0} to implement our models and baselines. We followed the original settings and tuned the parameters for each model. 
The embedding size was 64 for all models. We used Adam optimizer \cite{Adam} for most models except HGCFs/HRCFs/HICFs which used RSGD. 
The gate parameters of HGCH+ and the parameters of other models are initialized using Xavier initializer \cite{Xavier} except for the embedding weight of HGCFs/HRCFs/HICFs and HGCHs, which use uniform initializer and power law prior-based initializer, respectively. For NGCF, LightGCN, HGCFs, and HGCHs, we set their GCN layers to 3.\footnote{Although the HGCF reported the optimal layer number as 4, it ended up choosing 3 because they were almost identical.} We apply a grid search to the hyperparameters: the batch size of the different models on each dataset is tuned in $\{$1024, 2048, 4096, 8192$\}$ and the learning rate is tuned amongst $\{$0.01, 0.005, 0.001, 0.0005, 0.0001$\}$, weight decay or $L_2$ normalization is search in $\{$0.01, 0.005, $\cdots$, $5\times 10^{-5}$,$10^{-6}\}$, and the dropout ratio is tuned in $\{$0.0, 0.1, $\cdots$, 0.5$\}$ for NCF, NGCF. Besides, we employ the node dropout technique for NGCF, where the ratio is searched in $\{$0.0, 0.1, $\cdots$, 0.5$\}$. For WRMF, we select the positive item weight from $\{$1, 10, 100, 1000, 10000$\}$. For NCF, the ratio of positive and negative samples for negative sampling is 1:4, which follows the original setting. For DGCF, the iteration number and latent intents are set to 2 and 4, respectively, and the correlation weight is tuned amongst $\{$0.005,0.01,0.02,0.05$\}$. For HRCF, the $\lambda$ in the loss function is in the range of \{10, 15, 20, 25, 30\} and the layers of GCN is searched from 2 to 10.\footnote{In practice, we found that if training epochs are enough, the effect of the layer number is negligible.} For HICF, the margin is selected from $\{0.4, 0.5, \cdots, 1.0\}$, the $n_\text{neg}$ is tuned in $\{5, 10, 20, \cdots, 100\}$, and the layers of GCN is searched from 2 to 6. For our model, we set all curvatures to 1 for simplicity, leaving the exploration of different hyperbolic subspace curvatures with possible performance gains to future work. The hyperparameter of power law distribution $b$ and auxiliary task weight $\alpha$ are empirically set to 1.1 and 0.01 respectively. And the hyperbolic negative sampling number $n_\text{neg}$ is adjusted from 10 to 500. 
Compared to previous work that generally trained for 500 epochs \cite{HGCF,HRCF,HICF}, we extended the training to 1000 epochs because we found many methods had not yet converged. Additionally, consistent with prior work, we employed an early stopping strategy, meaning that training was stopped early if the NDCG@10 on the validation set did not improve for 100 consecutive epochs.

\section{In-depth Ablation Study}
In order to gain a deeper understanding of each component, we conducted a further analysis of the performance of different components on head items and tail items separately. In line with the work \cite{HICF}, the head and tail items are designated according to the 20/80 rule, which stipulates that items are ranked based on their degrees. The top 20\% are designated as the head (denoted H20), whereas the remaining 80\% are categorized as the tail (denoted T20). Tables \ref{tab:abla2} present the experimental results.
Here, HS symbolizes our Hyperbolic negative Sampling, PI stands for Power law prior-based Initialization, HA represents Hyperbolic neighbor Aggregation, and GP signifies Gate fusion with Prior. Parenthetical values reflect the improvement relative to the first row of the table.
From the results, we have the following insights:
\begin{itemize}[leftmargin=*]
\item The performance of HS for head items and tail items is not consistent across datasets (row 2 in Table \ref{tab:abla2}), indicating that HS may have different effects on different types of items depending on the dataset characteristics. On the other hand, PI+HA generally shows a preference for improving tail items (row 3 in Table \ref{tab:abla2}), suggesting that PI+HA can capture the long-tail distribution of items more effectively.
\item PI+HA+HS (\textit{i.e.}, HGCH, row 6 in Table \ref{tab:abla2}) and PI+HA+HS+GP (\textit{i.e.}, HGCH+, row 7 in Table \ref{tab:abla2}) consistently achieve better performance for tail items across all datasets, demonstrating their ability to handle the long-tail problem in recommender systems. Notably, Amazon-Book obtains an impressive maximum improvement rate of 68.1\% for tail items, which is much higher than the other datasets. This indicates that the hyperbolic space's exponentially expanding capacity enables the model to focus more on tail items \cite{HICF}, while our proposals emphasize this characteristic further by refining the hyperbolic model's structure and training.
\end{itemize}

\end{document}